\documentclass[letterpaper,journal,twoside]{IEEEtran}

\IEEEoverridecommandlockouts
\usepackage{cite}
\usepackage{amsmath,amssymb,amsfonts}
\usepackage{algorithmic}
\usepackage{graphicx}
\usepackage{textcomp}
\usepackage{xcolor}
\usepackage{stfloats}
\usepackage{comment}
\usepackage{subcaption}
\usepackage{booktabs}
\usepackage{threeparttable}
\usepackage{xfrac}
\usepackage{comment}
\usepackage{hyperref}
\usepackage{tabto}
\usepackage{orcidlink}
\usepackage{textcomp}
\usepackage{makecell}
\usepackage{array}
\usepackage{multirow}
\usepackage{enumitem}
\usepackage{xcolor}
\usepackage{soul}

\hypersetup{
    colorlinks=true,
    linkcolor=black,
    filecolor=black,      
    urlcolor=black,
    citecolor=black,
    }

\begin{document}

\title{A Comprehensive Design Framework for Vertical Power Delivery in High-Performance Computing}

\author{
Sriharini~Krishnakumar \orcidlink{0009-0009-6339-8718},
Yaroslav~Popryho \orcidlink{0009-0001-3382-882X}, Mingeun~Choi \orcidlink{0009-0009-5721-0678}, Ramin~Rahimzadeh~Khorasani \orcidlink{0000-0001-7879-7548},\\
Madhavan~Swaminathan \orcidlink{0000-0003-1729-2807}, Satish~Kumar \orcidlink{0000-0001-9701-8378} and
Inna~Partin-Vaisband \orcidlink{0000-0002-6399-6672} \vspace{-20pt}%

\thanks{Manuscript received 09 May 2026; revised 31 May 2026; accepted 00 March 2025. Date of publication 00 April 2025; date of current version 00 April 2025. This work was supported by the Center for Heterogeneous Integration of Micro Electronic Systems (CHIMES), one of seven centers in the Joint University Microelectronics Program (JUMP) 2.0, led by the Semiconductor Research Corporation (SRC) and sponsored by the Defense Advance Research Project Agency (DARPA).}%
\thanks{S. Krishnakumar, Y. Popryho, and I. Partin-Vaisband are with the Department of Electrical and Computer Engineering, University of Illinois Chicago, Chicago, IL 60607 USA (e-mail: skrish47@uic.edu;
ypopry2@uic.edu;
vaisband@uic.edu).
\textit{Corresponding author: Inna Partin-Vaisband.}}
\thanks{M. Choi and S. Kumar are with the George W. Woodruff School of Mechanical Engineering, Georgia Institute of Technology, Atlanta, GA 30332 USA (e-mail: mingeun.choi@gatech.edu; satish.kumar@me.gatech.edu).}%
\thanks{R. R. Khorasani and M. Swaminathan are with the School of Electrical Engineering and Computer Science, Pennsylvania State University, State College, PA 16801 USA (e-mail: ramin.rahimzadeh@psu.edu; mvs7249@psu.edu).}%
\thanks{Color versions of one or more figures in this article are available at
\url{https://doi.org/10.1109/TCPMT.2025.0000000}.}% <-this % stops a space
}

% \author{\IEEEauthorblockN{Sriharini Krishnakumar\IEEEauthorrefmark{1}, Yaroslav Popryho\IEEEauthorrefmark{1}, Mingeun Choi\IEEEauthorrefmark{2}, Ramin Rahimzadeh Khorasani\IEEEauthorrefmark{3},\\Madhavan Swaminathan\IEEEauthorrefmark{3}, Satish Kumar\IEEEauthorrefmark{2}, Inna Partin-Vaisband\IEEEauthorrefmark{1}}\\
% \IEEEauthorblockA{\IEEEauthorrefmark{1}{University of Illinois Chicago, IL, USA},\\\IEEEauthorrefmark{2}{Georgia Institute of Technology, GA, USA},\\ \IEEEauthorrefmark{3}{Penn State University, PA, USA}\\}
% Email: skrish47@uic.edu}
\maketitle

\begin{abstract}
Power delivery---including high-to-low voltage conversion, complex power distribution across heterogeneously integrated chiplets, and efficient interconnect allocation---remains a critical bottleneck in high-performance computing (HPC) systems. Existing vertical power delivery (VPD) solutions are estimated to achieve less than 70\% system-wide end-to-end power delivery efficiency, defined from platform input power to delivered on-chip load power, with substantial energy lost as heat before reaching on-chip point-of-loads (POLs). In the absence of systematic design methodologies, evaluating power quality, exploring architectural alternatives, and optimizing performance rely on computationally prohibitive simulations, resulting in suboptimal designs. This paper introduces an end-to-end scalable power delivery framework for HPC systems, including distributed VPD (DVPD) architecture, DVPD design optimization methodology, and analytical models. The framework leverages substrate-embedded GaN power switches together with arrays of unit inductors and capacitors tailored for HPC applications. Multi-stage power conversion schemes (48V-to-1V, 48V-to-24V-to-1V, and 48V-to-12V-to-1V) are explored, with system-wide voltage drops and power losses evaluated under 
steady-state conditions. Design specifications for passive and active devices are formulated to meet next-generation efficiency targets. For the 48V-to-1V case, the proposed DVPD approach achieves 84\% system-wide efficiency while occupying 54\% of the area beneath the load system, with efficiency increasing to 87.6\% at 75\% area utilization across a 1–50,kW load range. Furthermore, steady-state voltage drops peak at ~2.7\% and transient drops at 9\% (without decoupling capacitors), demonstrating the viability of DVPD for future wafer-scale HPC platforms.

\end{abstract}


\section{Introduction}
% --- Background ---
%Over the past two decades, rapid advancements in artificial intelligence (AI) have increased the demand for high-power, high-current-density processors, accelerators, and network devices essential for processing the growing size of data sets and machine learning (ML) models. 
%State-of-the-art (SOTA) power delivery solutions have already scaled to 1kW for individual chips, %(e.g., NVDIA V100) 
%and up to 10kW for server systems, %(e.g., NVIDIA DGX-V100), 
%yet system-level efficiencies remain below 70\% at current densities under 1~A/mm\textsuperscript{2} \cite{krishnakumar2023vertical}.
%According to the Heterogeneous Integration Roadmap (HIR), 
Within the next decade, the power demand of high-performance computing (HPC) systems is projected to increase to 1--2\,kW per chip and 20--50\,kW per server, with current densities reaching 2--4\,A/mm\textsuperscript{2}, according to the Heterogeneous Integration Roadmap (HIR)\,\cite{HIR2024}). However, the performance of representative power delivery solutions (e.g., 1\,kW NVIDIA V100, 10\,kW NVIDIA DGX-V100) remains limited, with system-wide efficiencies below 70\% and current densities under 1\,A/mm\textsuperscript{2}~\cite{khairy2020,
krishnakumar2023vertical}. This end-to-end efficiency is defined as the ratio of delivered chip power (TDP) to total platform input power, encompassing VR conversion, package PDN conduction, and platform distribution losses. The efficiency accounting methodology used throughout this work is formally summarized in Section IV.

% \begin{figure}[t!]
% \centerline{\includegraphics[width=0.5\textwidth]{figures/LPD.png}}
% \caption{Conventional lateral power delivery (LPD) architecture.}
% \label{fig:Conventional power architectures}
% \vspace{-15pt}
% \end{figure}

\begin{figure}[t!]
\centerline{\includegraphics[width=0.4\textwidth]{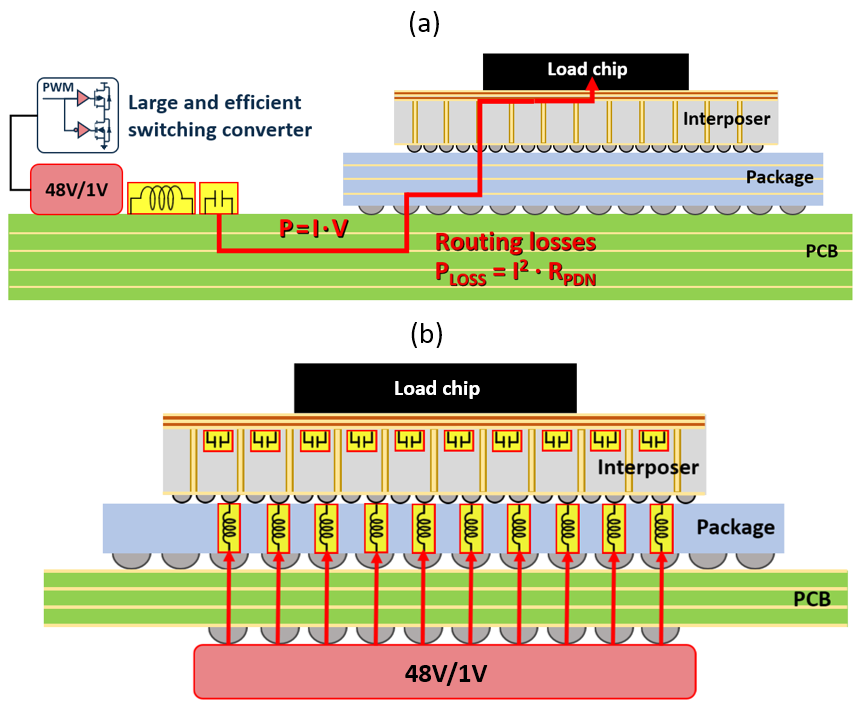}}
\caption{Power delivery approaches, (a) traditional lateral power delivery (LPD), and (b) vertical power delivery, featuring embedded passives and in-package power transistors integrated below the functional load chip.}
\label{fig: Conventional power architectures}
\end{figure}

%With advances in CMOS scaling and 3D/2.5D heterogeneous integration (HI) technologies, transistors now operate at ultra-low supply voltages ($<0.8$V) while drawing thousands of amperes. 
Traditionally, low-current 48\,V power is converted to high-current, 1\,V power with efficient but large motherboard voltage regulators (MBVRs). The converted high-current power is delivered via package power delivery network (PDN) to on-chip points-of-load (POLs) (see Fig.~\ref{fig: Conventional power architectures}a). In modern scaled-down and scaled-out, heterogeneously integrated systems, such package PDNs can dissipate up to $\sim$40\% of the total load power as heat~\cite{radhakrishnan2021power,Intel2024}. 

Furthermore, current density of existing passive and active power devices (e.g., inductors, capacitors, semiconductor switches) and the limited interconnect density constrain achievable power density in current architectures (e.g.,0.8\,A/mm\textsuperscript{2} for the NVIDIA H100\cite{choquette2023nvidia}). Consequently, novel power delivery architectures and power management techniques are required to address the escalating power demands of emerging HPC systems. 

Vertical power delivery (VPD) (see Fig.~\ref{fig: Conventional power architectures}b), incorporating compact high-ratio (48V/1V or 12V/1V) integrated voltage regulators (VRs), have emerged as a promising solution to mitigate traditional PPDN routing losses by reducing the effective VR-to-POL distance~\cite{geng2025vertically,kong2025vertical,liu2025monolithic,safari2023robust,safari2021power}. In this approach, low current is distributed laterally at high voltage prior to conversion, and converted with high-ratio VRs near POLs. As a result, the converted high current is only delivered vertically over a short distance to the on-chip loads. Compared to the MBVR-based 48\,V/1\,V conversion, this approach can reduce routing losses by nearly $48^{2}\times$. %compared to conventional lateral power delivery (LPD) architectures, depending on whether the IVRs convert directly from 48V or from 12V, respectively. 
For example, less than 5\% routing loss has been reported for a 1\,kW VPD system in~\cite{krishnakumar2024design}. In addition, vertical stacking of passive and active power components in VPD architectures enable higher system-wide power density. 

%However, VPD architectures face inherent challenges: (1) stringent in-substrate area constraints necessitate miniaturized passive and active devices, which require high switching frequencies (1–10\,MHz) and thereby increase switching losses, and (2) reliance on a single or few VRs substantially elevates internal VR conduction losses, which scale quadratically with the output current and are further aggravated by the higher DC resistance of miniaturized devices. Hence, while traditional power architectures suffer from high routing losses, VPD architectures incur high conversion losses. 

Despite these advantages, conventional lumped VPD architectures face fundamental limitations. Mechanically, embedding a single, large, stiff converter within a substrate produces significant stiffness gradients and coefficient of thermal expansion (CTE) mismatch with respect to the surrounding substrate, inducing global package warpage, delamination risk, and shear stress in the redistribution layer (RDL). From a reliability standpoint, strong electrical coupling between the converter power stage, the PDN, and the load system permits faults within the VR to propagate through the shared PDN within sub-microsecond timescales, risking system-level failure and making fault containment a primary reliability concern.  Furthermore, a monolithic VR limits diagnostic resolution and prevents localized fault isolation within the PDN. Pre-assembly characterization and post-assembly fault isolation are also limited, as individual converter stages cannot be independently accessed once integrated within a monolithic embedded power-delivery structure. Electrically, although lumped VPD architectures reduce routing loss, substantial voltage drop persists within the horizontal interconnects of the RDLs, as current must travel long distances—up to a quarter of the system perimeter—between the VR output and distant POLs on the same or adjacent chiplets.
Finally, a centralized VR cannot manage spatio-temporal variations in HPC system power demand, as load steps through the shared PDN generate large current transients, increasing voltage deviation and electrical stress on the converter and global decoupling network.

To address these limitations, a distributed vertical power delivery (DVPD) is proposed. The architecture employs a fine-pitch network of horizontally distributed VRs, with VR components vertically stacked across multiple power layers (PL), placing the VR outputs' just tens of microns from the on-chip POLs. Partitioning power conversion into multiple smaller VRs reduces peak stress concentration and prevents the formation of a single mechanically stiff region within the substrate. Each VR produces localized deformation rather than global package deflection. With symmetric VR placement localized mechanical effects can partially offset one another, reducing overall warpage sensitivity. The distributed architecture also provides inherent redundancy. Real-time fault detection and isolation enable selective deactivation of a faulty VR while the remaining regulators redistribute load current. System operation therefore continues under partial failure conditions, improving fault tolerance and mitigating risk of system-level failure \cite{krishnakumar2026power}. Moreover, the modular and identical VR-cell structure of DVPD enables cell-level characterization prior to assembly, and the selective VR activation capability that supports fault tolerance equally provides a mechanism for in-system functional testing. A complete embedded-array test insertion methodology remains an important direction for future work of \cite{krishnakumar2026power}.

%From efficiency standpoint a DVPD system with $M$ VRs, the per-VR output current is reduced to ${I_{o}}/{M}$, compared to ${I_{o}}$ in a single-VR approach. Consequently, the internal VR conduction loss and thermal dissipation in power components (e.g., power switches) scale as $M(I_{o}/M)^{2}=I_{o}^{2}/M \ll I_{o}^{2}$, thereby mitigating the high conversion losses inherent in lumped VPD architectures. Although miniaturization of VR components may increase their DC resistance, this increase scales \texttt{sublinearly} with $M$ (approximately as $\sqrt{M}$, as shown in Section III.C) and does not offset the substantial reduction in VR conduction loss. In addition as $M$ increases, power components move closer laterally (with constant trace width; see Section III.D), further reducing conduction losses in RDL. Hence, the distributed architecture effectively lowers overal power conversion losses, while the achievable vertical stacking serves (applicable to both DVPD and VPD) primarily to enhance system-wide power density.
From an efficiency standpoint, in a DVPD system with $M$ VRs, the per-VR output current is reduced to $I_o/M$, compared to $I_o$ in a single-VR approach. As $M$ increases, the VR components move closer laterally, shortening the intra-VR routing paths and reducing conduction losses in the RDL network (with constant trace width, see Section~\ref{subsec:Sizing}). Hence, the distributed architecture effectively lowers internal routing losses within VR, while vertical stacking (applicable to both DVPD and VPD) primarily enhances system-wide power density.  Although VPD can increase density, its single-inductor high-current requirement typically forces low-density air-core inductors, whereas DVPD lowers per-inductor current and enables dense magnetic-core integration. 

Moreover, DVPD architectures inherently support dynamic, fine-grained power management—an increasingly critical capability for heterogeneous, chiplet-based next-generation HPC systems \cite{krishnakumar2026dynamic}.

It is important to distinguish the proposed DVPD approach from conventional board-level distributed VRM placement. Existing systems may employ multiple VRMs on the board or package periphery; however, these converters are typically few in number, relatively coarse-grained, and still deliver high current laterally through the package PDN before reaching the on-chip POLs. In contrast, DVPD considers a much more highly distributed in-package design regime, where tens to hundreds of vertically integrated VR cells are embedded beneath the load system and co-optimized with the package PDN. At this scale, power delivery becomes a coupled design problem involving converter topology, switching frequency, passive-component scaling, floorplanning, vertical interconnect allocation, routing loss, ripple constraints, and package-level area limits. Therefore, practical DVPD requires not only an architectural concept, but also a systematic design methodology and automated framework for rapid performance evaluation, design-space exploration, and optimization.

Existing methodologies rely on computationally expensive simulations, rendering system-level evaluation and optimization impractical for large-scale power architectures such as the proposed DVPD. The proposed framework provides a scalable DVPD design methodology spanning from individual high-performance chips operating at a few kilowatts to wafer-scale systems operating at tens of kilowatts. Rather than assuming a fixed number of converters or a manually selected layout, the framework determines the degree of VR distribution, component sizing, placement, and vertical stacking required for a given power specification and set of technology constraints, providing quantitative estimates of power-loss distribution, DC voltage drops, and transient voltage droops across large-scale DVPD systems. The primary contributions and new findings of this work are as follows:
%The proposed framework presents a scalable DVPD design optimization methodology, spanning from high-performance individual chips operating at a few kilowatts to wafer-scale systems operating at tens of kilowatts. 
%The proposed framework determines the degree of distribution and vertical stacking required in a DVPD system for given power specifications and in doing so provides several new insights. Specifically, it reveals: (1) the optimal number and sizing of VRs required to maximize conversion efficiency under area constraints, (2) the VR placement coordinates that minimize PPDN routing losses, (3) the optimal number of vertically stacked layers needed to maximize power density, and (4) quantitative estimates of power loss distribution, DC voltage drops, and transient voltage droops across large-scale DVPD systems.
\begin{itemize}[leftmargin=10pt]
    \item \textbf{DVPD architecture:}
    A DVPD architecture is introduced in which power conversion is spatially partitioned across VR tiles with vertically integrated passive and active components. By restructuring current distribution in both lateral and vertical dimensions, DVPD mitigates copper-density concentration, stiffness gradients, and CTE mismatch, enabling dynamic power management and fault-tolerant scalable multi-kilowatt power delivery.
    
    \item \textbf{Loss-optimized VR partitioning framework:}
    A quantitative VR sizing methodology is developed that balances conduction and switching losses under parasitic constraints. Applicable to both VPD and DVPD, the formulation reveals sublinear scaling of dominant losses with regulator partitioning under optimal switch sizing, providing a principled basis for converter-level scaling.
    
    \item \textbf{Unified DVPD design methodology and system-wide co-optimization:}
    A system-level framework jointly optimizes VR count, sizing, placement, vertical stacking, and interconnect allocation under packaging constraints, explicitly coupling converter design, floorplanning, and PDN behavior for physically consistent design-space exploration.
    
    \item \textbf{Scalable architectural evaluation:}
    The framework enables systematic comparison of optimized VPD and DVPD configurations across integration levels (48V/1V, 48V/24V/1V, 48V/12V/1V), demonstrating scalable operation from 1~kW to tens of kW and achieving up to 84\% system-wide efficiency within the modeled design space.

\end{itemize}
%the following output 
% \begin{enumerate}
%     \item Optimal number and sizing of VRs for maximized conversion efficiency while satisfying area constraints,
%     \item Optimal VR location coordinates for minimized PPDN routing losses,
%     \item Optimal number of vertically stacked layers for maximized power density, and
%     \item Estimated of power loss distribution, DC voltage drops, and transient voltage droops.
% \end{enumerate}

%
\begin{figure*}[t!]
\centerline{\includegraphics[width=\textwidth]{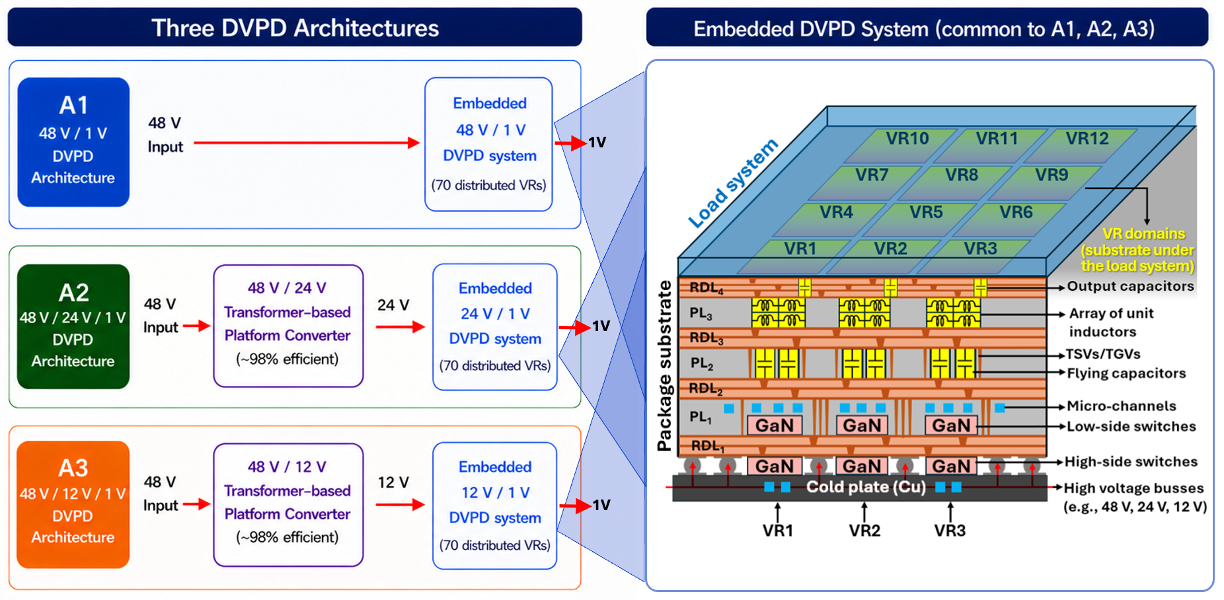}}%{figures/DVPD1.png}}
\caption{ Architecture A1, A2, A3 employ DVPD architecture with power transistors and passives embedded in-substrate below the die for 48~V/1~V, 24~V/1~V, and 12~V/1~V, respectively. In addition, for A2 and A3, a first-stage conversion (48~V/24~V and 48~V/12~V) is performed on PCB using high-efficiency ($\sim$98\%) transformer-based converters.}
\label{fig:DVPD}
\end{figure*}

The proposed framework enables rapid construction and evaluation of practical power delivery solutions for scaled-out systems while rigorously capturing the effects of device- and package-level technologies as well as alternative power conversion circuit implementations. It has also been extended to thermal-VPD co-design studies~\cite{choi2025substrate, choi2024thermal} and in ongoing investigations of HPC cost and performance tradeoffs. To ensure realistic behavior, the framework is built upon validated device, converter, and interconnect models cross-verified against manufacturer tools, prior implementations, SPICE/Spectre simulations, and extracted parasitics, with technology parameters derived from datasheets, foundry design rules, and prior experimental studies.

The remainder of this paper is organized as follows. Section~\ref{sec: VPD} describes the DVPD architecture featuring vertically stacked in-substrate and backside VR components. In Section~\ref{sec:Methodology}, a performance-driven design optimization methodology for HPC DVPD is introduced. Analytical models for determining the minimum switching frequency, optimal sizing of power switches, and vertical interconnect allocation are also presented in this section. In Section \ref{sec:Results}, the framework is exploited to design, characterize, and compare different power delivery architectures. Finally, the paper is concluded in Section~\ref{sec:Conclusion}.

\section{Distributed Vertical Power Delivery (DVPD)}\label{sec: VPD}

The proposed DVPD architecture, supporting both single-stage (48V/1V) and dual-stage (48V/24V/1V or 48V/12V/1V) power conversion schemes, is illustrated in Fig.~\ref{fig:DVPD}. The architecture comprises $M$ laterally distributed VRs, with the corresponding power devices vertically aligned and embedded within stacked substrate layers. In the single-stage configuration, the 48~V low-current supply is delivered to the individual VRs at the bottom substrate layer and flows vertically through the stack while being converted to 1\,V high-current POL power. In the dual-stage configuration, the first conversion stage (48V/24V or 48V/12V) is implemented on the system integration platform (e.g., PCB, wafer, panel), and the intermediate 24\,V or 12\,V power is subsequently converted and delivered to POL by the distributed VRs with vertically stacked components.

Multi-phase hybrid converters (switched-capacitor (SC) buck) are widely recognized for their high efficiency in high-ratio voltage conversion, as well as for alleviated voltage stress on power switches and effective mitigation of the ultra-low duty cycle inherent to buck converter \cite{kirshenboim2017high-efficiency,das2019regulated,gong202290,baek2021vertical,pillonnet2024analytical,zhu2021multi}. 
Furthermore, multi-phase VR topologies, consisting of several parallel-connected, phase-shifted branches, are commonly employed to reduce output current ripple \cite{parisi2017multiphase}. In this work, a phase refers to a switching branch (comprising an inductor and its corresponding power switches) that operates with a phase shift relative to other branches, delivering a portion of the total converter current. Accordingly, per-phase metrics refer to the performance of one such branch.
%Each branch—comprising an inductor and its corresponding power switches—delivers a phase-shifted portion of the total converter current. Throughout this paper, per-phase metrics refer to the performance of a single branch within a multi-phase (or single-phase) VR.

%
\begin{figure*}[t!]
\centerline{\includegraphics[width=\textwidth]{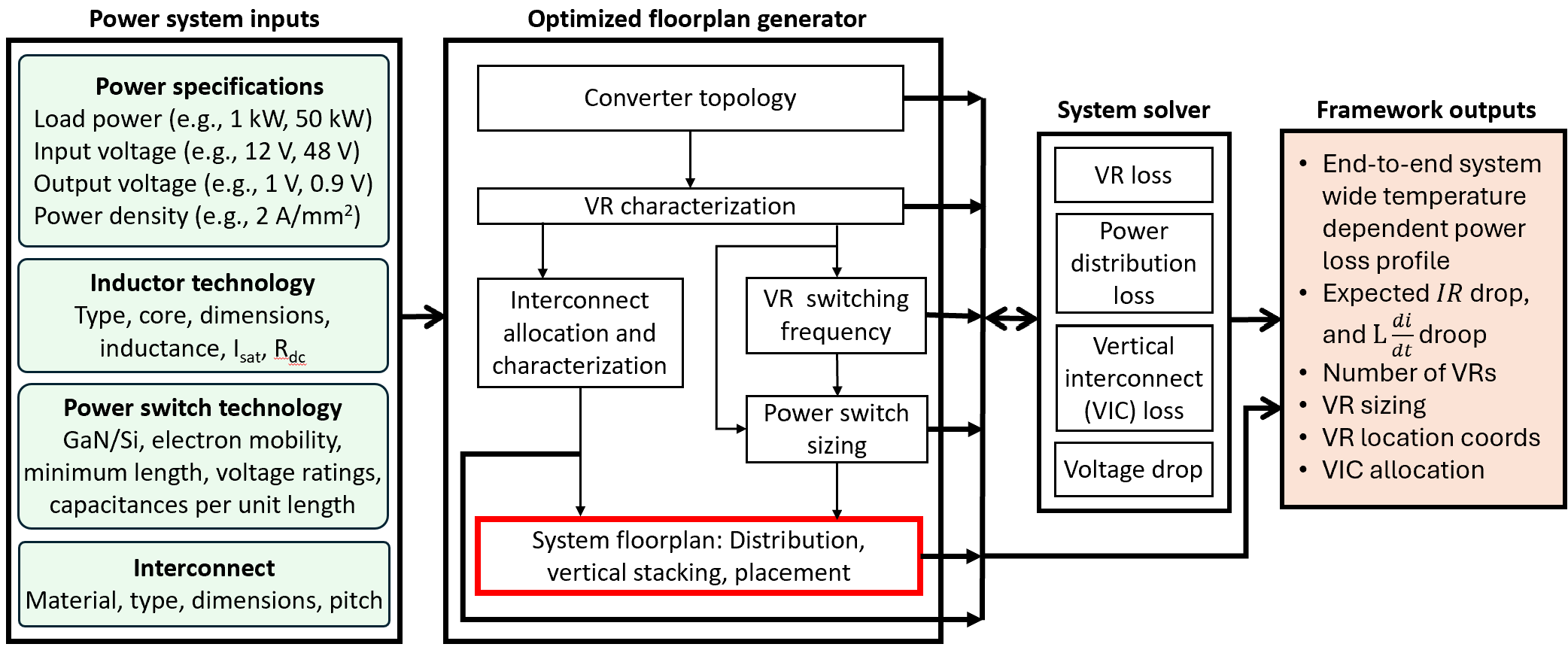}}
\caption{Proposed design optimization framework for DVPD systems. Design components are described in Section \ref{sec:Methodology}.}
\label{fig:Framework}
\end{figure*}

In the proposed DVPD architecture, these converters are particularly advantageous, as their multi-phase operation enables efficient current sharing among distributed VRs, while their hybrid topology supports high voltage conversion ratios, power density and compatibility with vertically stacked integration. The proposed framework is not limited to a specific hybrid converter implementation; rather, it provides a general design methodology for co-optimizing converter components (high- and low-side power switches, SC flying capacitors, and buck inductors) within vertical stacks of DVPD power layers. Furthermore, the proposed framework is equally applicable to other switching-based topologies that exploit high-side and low-side switches, ranging from traditional buck converters to emerging resonant VRs~\cite{lotfi2026interleaved}. %Hence in this work hybrid converer topologies are considered for 48V/1V conversion single-stage architecture, as well as for second-stage conversion (24V/1V or 12V/1V) in dual-stage architectures. Such topologies incorporate high-side switches and flying capacitors forming a switched capacitor (SC) sub-circuit to step down the high input voltage to an intermediate level, which is further reduced to the POL voltage by low-side switches and inductors, similar to a buck topology. 

%Based on prior studies~\cite{krishnakumar2023vertical,abdelzaher2025hybrid}, the double series capacitor hybrid (DSCH) converter is benchmarked in this work for 48V/1V single-stage conversion, as well as for second-stage conversion (24V/1V or 12V/1V) in dual-stage architectures, owing to its high efficiency and use of fewer components, which enables higher power density. 
%hybrid configurations (e.g., DSCH converter, as shown in Fig. \ref{fig:DSCH}) incorporate high-side switches ($Q_c$, $Q_{1a}$, $Q_{1b}$ in Fig. \ref{fig:DSCH}) and flying capacitors (e.g., $C_{t1}$, $C_{t2}$ in Fig. \ref{fig:DSCH}) forming a switched capacitor (SC) sub-circuit to step down the high input voltage to an intermediate level, which is further reduced to the POL voltage by low-side switches ($Q_{2a}$, $Q_{2b}$ in Fig. \ref{fig:DSCH}) and inductors ($L_a$, $L_b$ in Fig. \ref{fig:DSCH}), similar to a buck topology. 

\begin{figure}[t!]
\centerline{\includegraphics[width=\linewidth]{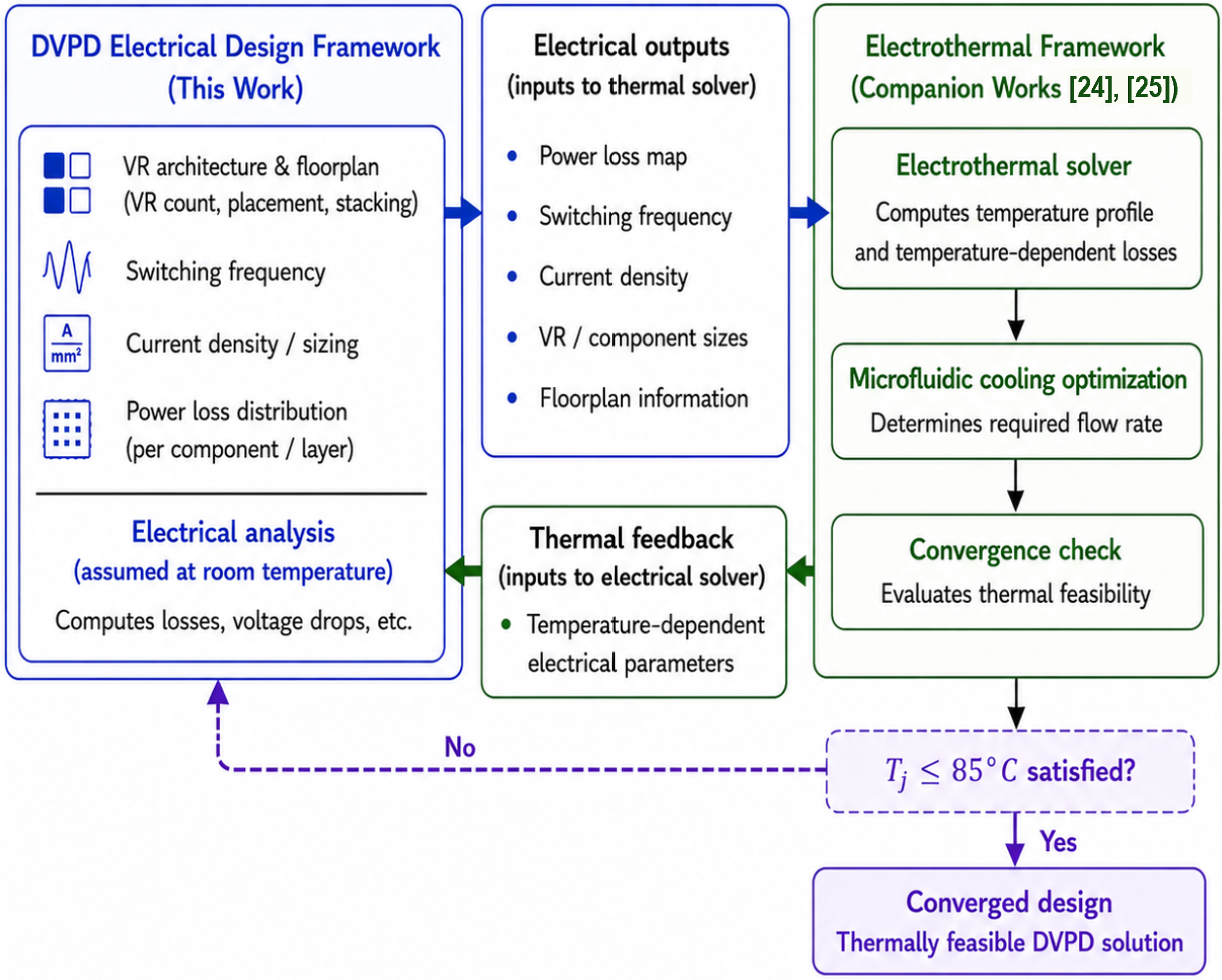}}
\caption{Interaction between the proposed DVPD electrical design framework and the self-consistent electrothermal framework of~\cite{choi2025automated,choi2025self}.} 

%\hl{Electrical and architectural outputs from the DVPD framework are supplied to the electrothermal solver, which dynamically adjusts microfluidic cooling conditions to maintain thermal constraints and evaluates temperature feasibility under realistic operating conditions.}

\label{fig:em-frame}

\end{figure}

Hybrid VR components are integrated within multiple vertically stacked substrates  (e.g., three in Fig.~\ref{fig:DVPD}), aligned with the natural power flow of the hybrid topology. Furthermore, when fabricated as individual chiplets, VR components can be vertically stacked on the backside of a substrate and interconnected through TSVs. %In this paper, each embedded or backside layer of VR components is designated as a power layer (PL). 
An example of the multi-substrate approach in Fig.~
\ref{fig:DVPD} employs high-side GaN switches that are flip-chipped to the RDL of the bottom substrate, while low-side switches (bottom layer), flying capacitors (middle layer), and inductors (top layer near the POLs) are embedded within the subsequent substrate power layers. To guarantee scalability of the architecture to high-power scale-out systems, the power delivery system must fit within the footprint of the load. Accordingly, the number of power layers is ultimately determined by the number and size of components in the chosen VR topology and constrained by the capabilities of advanced packaging technologies. This establishes a direct link between VR topology selection and practical packaging limits, highlighting a critical design tradeoff for scalable DVPD systems. %For example, a simple multi-phase buck converter can be designed with only two substrate layers (one for low-side switches and one for inductors) and high-side switches flip-chipped to the backside of the substrate. 

In the VRs of interest, more than 70\% of conversion losses originate from power switches, making them the primary source of heat dissipation, with temperatures exceeding 100\textdegree C \cite{choi2024thermal,choi2025substrate}. This necessitates advanced thermal management techniques, such as substrate-embedded micro-channel cooling to effectively dissipate heat and maintain power system performance. In the proposed framework, thermal awareness is incorporated using the thermal-fluidic solver of~\cite{choi2025automated,choi2025self} enabling evaluation of cooling requirements and temperature feasibility during architectural DVPD optimization. Specifically, the component-level power-loss distributions and optimized DVPD geometry generated by the proposed methodology are supplied to the thermal framework, which iteratively updates temperature-dependent losses and cooling conditions until thermal convergence is achieved as shown in Fig. \ref{fig:em-frame}. Consequently, density-efficiency tradeoffs are evaluated under thermal feasibility constraints.

Embedding chiplets into substrates with advanced packaging has recently gained attention in applications such as mm-wave telecommunication \cite{watanabe2020review} and power delivery \cite{khorasani2025embedded,deprospo2024saras}, owing to reduced interconnection losses enabled by shorter via lengths. Recent studies have explored various substrate materials and fabrication processes, with glass emerging as a promising candidate for embedding single or multiple chiplets \cite{li2024glass}. However, while chip embedding within a single substrate layer has been demonstrated, the fabrication of vertically stacked substrate layers that integrate both active and passive components remains an open challenge. Based on the findings of this work, advances in multi-substrate vertical stacking and chiplet embedding are identified as critical enablers for meeting the power performance goals defined in the HIR~\cite{HIR2024}.

\section{Co-Optimization of VRs and PPDN for DVPD}\label{sec:Methodology}

\begin{comment}
\begin{figure}[b!]
%\vspace{-5pt}
\centerline{\includegraphics[width=0.5\textwidth]{figures/DSCH.png}}
\caption{Double series capacitor hybrid (DSCH) converter.}
\label{fig:DSCH}
\end{figure}
\end{comment}
%
The design of complex DVPD systems requires careful co-optimization of the multi-layer PPDN and distributed stacks of VRs, while accounting for the physical, electrical, and thermal properties of the VRs in conjunction with advanced packaging and device constraints. All DVPD components, including horizontal RDLs, vertical interconnects, passive devices, and power switches, must be jointly co-optimized to maximize overall system efficiency. 

%A unified design framework for DVPD systems is presented in Fig.~\ref{fig:Framework}, in which switching frequency selection, power switch and inductor sizing, interconnect allocation, and system floorplanning are co-optimized within a single flow. By integrating component libraries with analytical DVPD-aware loss and voltage models, end-to-end system outputs are obtained, including the number, size, placement, and vertical stacking of VRs, along with system-wide efficiency, IR drop and transient droop estimates. From the application of the framework, critical dependencies between converter topology, interconnect allocation, and packaging constraints are revealed, providing quantitative insights into achievable efficiency and power density under the packaging and device technology constraints. Through these findings, DVPD is established as a scalable and efficient solution, and practical design principles are derived that guide the realization of high-power, high-density computing systems.

A unified design framework for DVPD systems is presented in Fig.~\ref{fig:Framework}, in which switching frequency selection, power-switch and inductor sizing, interconnect allocation, and system floorplanning are co-optimized within a single flow. By combining component libraries with analytical DVPD-aware loss and voltage models, the framework determines the number, size, placement, and vertical stacking of VRs while evaluating system-wide efficiency, IR drop, and transient droop. The resulting design-space exploration reveals key dependencies among converter topology, packaging constraints, and interconnect allocation, providing quantitative insight into achievable efficiency and power density.

Building on prior studies~\cite{krishnakumar2023vertical,abdelzaher2025hybrid}, the double series capacitor hybrid (DSCH) converter %(see Fig. \ref{fig:DSCH})
is adopted in this work as a promising architecture for high-voltage ratio conversion (from 48\,V, 24\,V, or 12\,V to 1\,V), and, more importantly, as a demonstration vehicle to expose key concepts, dependencies, and design tradeoffs in DVPD. By applying the proposed framework to DSCH, findings related to operating frequency and VR sizing are derived, which generalize to other converter topologies. Specifically, the section is organized into five subsections: (a) VR switching frequency, (b) modular inductor scaling, (c) power switch sizing, (d) floorplanning, and (e) interconnect considerations.
%The framework is demonstrated based on a DVPD system, with up to 100 embedded DSCH VRs, delivering 1 kW of load power at 2 A/mm\textsuperscript{2} to a 500 mm\textsuperscript{2} system (e.g., a single functional die or collection of several chiplets closely integrated).

\subsection{VR Switching Frequency}\label{subsec:Frequency}

The switching frequency of hybrid regulators plays a critical role in system performance. 
%\hl{Higher switching frequencies reduce ripple but simultaneously increase switching losses, creating a fundamental tradeoff that must be balanced in the design process.} 
Higher switching frequencies reduce the required inductance and passive-component area but simultaneously increase switching-related losses and thermal density. Since the proposed framework minimizes total power loss subject to footprint, power-density, and ripple constraints, the optimal operating point for the evaluated DVPD configurations naturally occurs near the minimum feasible switching frequency satisfying the specified ripple constraint. To address this tradeoff in a systematic manner, a generalizable methodology is developed for selecting a preferred VR switching frequency. This methodology provides design guidelines that are applicable across hybrid converter topologies and packaging configurations, ensuring that ripple is maintained within acceptable limits while minimizing switching losses \cite{vaisband2017heterogeneous}. The approach is demonstrated based on a DSCH VR in Section \ref{sec:Results}.

\begin{figure}[t!]
%\vspace{-10pt}
  \centerline{\includegraphics[width=0.4\textwidth]{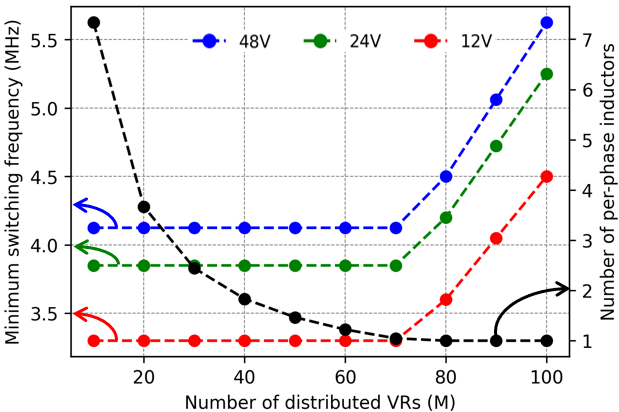}}
  \caption{Minimum switching frequency, \(f_{\text{sw}}^\text{min}\), of DSCH VRs in a DVPD system delivering 1kW@1V load power to the functional system at a current density of 2 A/mm\textsuperscript{2}.}
  \label{fig:Minimum switching frequency}
\end{figure}

Power inductors in multi-phase hybrid topologies, operating in continuous conduction mode, carry a DC-biased triangular ripple current. 
Each phase, consisting of an inductor and its switches, delivers a portion of the load current. The per-phase ripple amplitude ($\Delta i_\varphi = \gamma I_\varphi$) decreases with larger inductance ($L_\varphi$) and higher switching frequency ($f_{sw}$), and is constrained to maintain efficiency and device reliability. The minimum switching frequency therefore scales as,
\begin{equation}\label{eq:F_ind}
    f_{sw}^\text{min} \propto \frac{\Delta V_L}{\gamma I_\varphi L_\varphi},
\end{equation}
where $\Delta V_L$ is the effective voltage swing across the inductor (e.g., 48\,V$\to$1\,V or 12\,V$\to$1\,V conversion). 

Both $\Delta V_L$ and $L_\varphi$ are constrained in DVPD: the large step-down conversion produces a high inductor voltage swing, while embedded inductors necessarily provide smaller inductance. This creates a fundamental tradeoff: distributing current across more regulators 
%reduces conduction losses (${I^2R \;>\; (\frac{I}{M})^2MR}$), but also 
lowers the per-VR current (and potentially the conduction losses), but also proportionally lowers the permissible current ripple which increases switching losses by pushing $f_{sw}^\text{min}$ higher. \texttt{Multi-phase operation is essential} to alleviate this tradeoff, since interleaving phases relaxes the ripple constraint and reduces the minimum switching frequency, enabling higher efficiency and scalability in DVPD.

\subsection{Modular Inductor Scaling}

Inductors and power switches are the largest VR components, and one of them generally dictates the footprint within which all other devices must be accommodated on adjacent vertically stacked layers. In addition, the size of these devices directly impacts conduction and switching losses as well as thermal density of the stacked VR. 

A primary limitation of compact magnetic-core inductors is the saturation current. To sustain the target load current within the footprint constraints, a modular inductor approach is adopted, trading excess inductance for increased saturation capability. In this approach, multiple unit inductors are connected in parallel and arranged both laterally and vertically in a fine-pitch 3D array (see Fig.~\ref{fig:DVPD}). Each unit inductor exhibits an inductance $L_0$ and a saturation current $I_0$
. When the required per-phase inductor current ($I_{\phi}$) exceeds $I_{L0}$, several unit inductors are paralleled, with the required number given by ${N_{L} = \lceil I_{\phi} / I_{L0} \rceil}$. This configuration reduces the effective inductance to ${L_{\phi} = L_{0} / N_{L}}$, thereby increasing the minimum switching frequency as defined in~(\ref{eq:F_ind}). 
Area utilization is maximized when ${I_{\phi}/I_{L0}}$ is close to an integer, minimizing redundant units and maximizing current density within the footprint. Accordingly, \texttt{unit inductors should be designed for scalable parallel integration} by carefully balancing saturation current and inductance values. 
In addition, by adjusting the number of parallel-connected inductors, the effective inductance and current capacity can be tuned to match operating conditions (e.g., higher current with lower effective inductance, or lower current with higher inductance to meet ripple constraints). While power management is not explicitly modeled in this paper, it remains a primary concern and is expected to benefit from the modular inductor approach.

For such scalable design, the minimum switching frequency in~(\ref{eq:F_ind}) is expressed as
\begin{equation}\label{eq:F_ind0}
    f_{sw}^\text{min} \propto \frac{\Delta V_L}{\gamma I_\varphi L_\varphi} = 
    \frac{\Delta V_L}{\gamma I_{L0} L_0},
\end{equation}
where $\gamma$ is the allowable ripple ratio.

The relationship between minimum VR switching frequency, the number of distributed VRs, and the input voltage is illustrated in Fig.~\ref{fig:Minimum switching frequency}. Over a broad range of VR counts (up to $\sim$70 units in this example), the minimum frequency remains flat because allowable ripple ratio, $\gamma$, is achieved through the required number of unit inductors, $N_L$, independently of the per-VR current (see~(\ref{eq:F_ind0})). Once the per-VR current drops below the saturation rating of a single unit inductor and $N_L$ cannot be further reduced, the allowable ripple ratio must be achieved by driving the minimum switching frequency upward. Note that beyond this threshold, the inductor overhead increases

Input voltage further shifts this behavior: larger conversion ratios (e.g., 48\,V$\to$1\,V) impose a higher minimum frequency than smaller ratios (e.g., 12\,V$\to$1\,V). Together, these results indicate the existence of a preferred VR count that minimizes both per-phase inductor usage and switching frequency. While the exact crossover depends on system targets (here shown for a 1\,kW DVPD system at 2\,A/mm$^2$), the general insight is that the proposed framework identifies the regime where distributed VRs achieve the lowest \texttt{feasible switching frequency without excessive inductor overhead}.

\begin{figure}[t!]
\centerline{\includegraphics[width=0.4\textwidth]{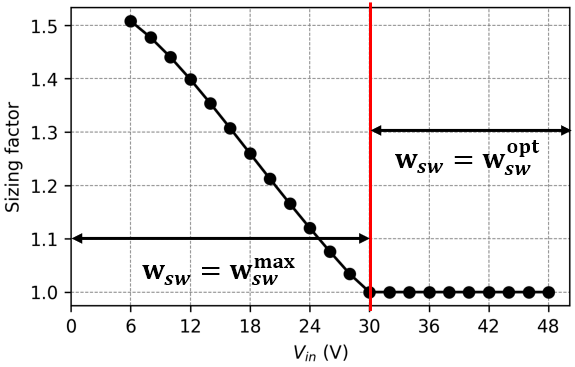}}
\caption{Sizing factor, ${w_{sw}^\text{opt}}$, in a DVPD system delivering 1\,kW load power. The figure illustrates the general trend that the sizing factor decreases with 
input voltage, with the crossover point indicating the voltage at which footprint constraints begin to limit switch sizing.}
\label{fig:Sizing factor}
\end{figure}

\subsection{Power Switch Sizing}\label{subsec: Power Switch Sizing}
In high-ratio embedded VRs, the dominant portion of power loss occurs in the %\textit{high-side} 
power switches. %(while the inductor loss typically constitutes a comparable or secondary component, as discussed in Section IV). 
Therefore, optimizing the power switches to minimize conduction and switching losses is essential for improving overall converter efficiency. %Alternatively, the VR footprint is dominated by the \textit{low-side} switches (along with the inductors), as they conduct the output current for a larger fraction of the switching cycle, resulting in higher RMS current and therefore requiring a larger device width. %Such decoupled efficiency and density enable an effective optimization of VR resources, as explained in this section.

%A typical configuration of a high-side (HS) and low-side (LS) power switches with associated parasitic capacitances is illustrated in Fig.~\ref{fig:...}. 
In the general case, the optimum width values of the individual VR power switches can be obtained by constructing the total power-loss expression \cite{krishnakumar2024vertical}: larger currents favor wider devices (with lower ON resistance $\propto 1/w$), while higher switching frequencies penalize wide devices due to increased capacitive losses ($\propto w \cdot f_{sw}$). Solving for ${\partial P/\partial w=0}$ the scaling of the optimal switch width as, 
\begin{equation}\label{eq:wopt-lin}
    %w_{HS}^{\mathrm{opt}}, \; w_{LS}^{\mathrm{opt}} \;
    w_\text{FET}^{\mathrm{opt}} \;\propto\; 
    \frac{I_{ds}}{V_{ds}} \cdot \frac{1}{\sqrt{f_{sw}}},
\end{equation}
where $I_{ds}$ and $V_{ds}$ are, respectively, the RMS current 
and off-state voltage across the power switch. For a given load current per VR, the optimal switch width follows~(3). 
With $M$ distributed VRs, the current handled by each VR scales as $I_{ds} \propto 1/M$, 
yielding $w_{\mathrm{FET}}^{\mathrm{opt}} \propto 1/M$. Consequently, the per-VR conduction 
and switching losses decrease approximately as $1/M$. Since $M$ VRs operate in parallel, the reduction in per-VR loss is offset by the number of VRs, such that the total power-switch loss remains approximately constant for a fixed system-level load power.

The \textit{low-side} switches conduct output current for a larger fraction of the switching cycle and therefore carry higher RMS current, resulting in larger device width than the high-side switches. Consequently, the overall VR footprint is primarily determined by the low-side switches and the embedded inductors.

The intent of (\ref{eq:wopt-lin}) is to capture the topology-independent scaling relationship of the optimal switch width with current, voltage stress, and switching frequency that drive the proposed design methodology. The exact switch widths are obtained as part of the optimization framework by constructing the total switch-loss expression for the selected device technology and converter topology and solving ${\partial P/\partial w=0}$. Representative VR power-loss formulations are provided in \cite{krishnakumar2024vertical}, while analogous models are commonly used for GaN HEMTs and other power-device technologies.

\subsection{Floorplanning}\label{subsec:Sizing}

\begin{figure}[t!]
    \centering
    %\vspace{-10pt}
    \includegraphics[width=\columnwidth]{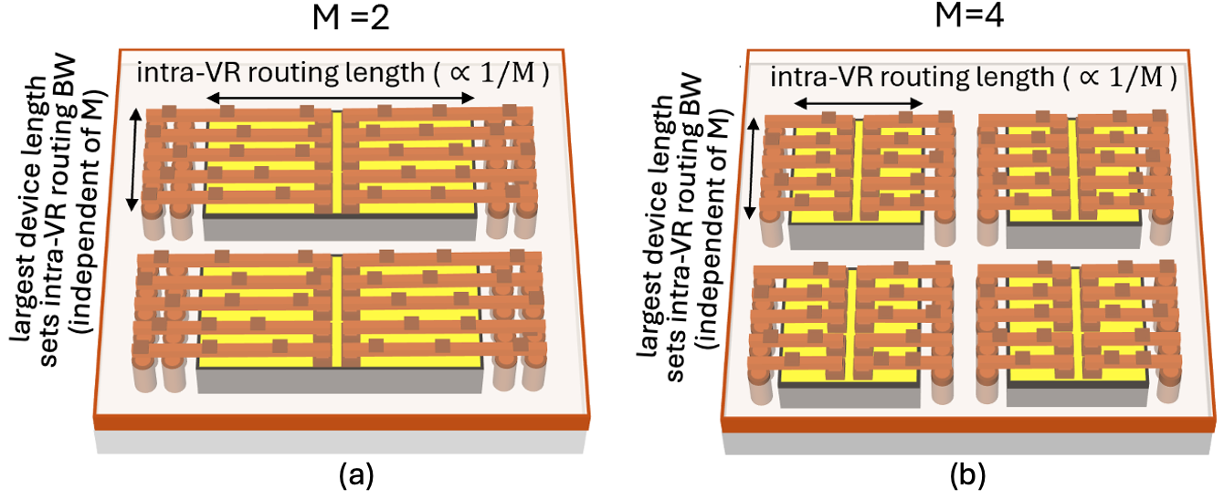}
    \caption{Intra-VR routing geometry scaling with increasing number of distributed voltage regulators (VRs). While the per-VR footprint decreases with increasing $M$, the effective routing width remains largely unchanged, whereas routing length decreases approximately with $w_\mathrm{LS}^\mathrm{opt} \propto 1/M$. Consequently, interconnect resistance and associated intra-VR routing losses are reduced, providing a key benefit of the proposed DVPD architecture.
    }
    \label{fig:Routing_length_scaling}
\end{figure}

Given $M$ VRs distributed across vertically stacked layers and a fixed per-layer area budget (e.g., 250\,mm\textsuperscript{2} if supplying 1\,kA at 4\,A/mm\textsuperscript{2}), the VR components on each layer (e.g., switches on a switch layer, or inductors on an inductor layer, as shown in Fig.~\ref{fig:DVPD}) are arranged in an $m \times n$ array. The number of rows $m$ in this array is determined by the dimensions of the largest components, typically the inductor length ($l_{ind}$) or the switch length ($l_{sw}$). Accordingly, the maximum allowable low-side switch width is given by
\begin{equation}\label{eq:W_max}
    w_{LS}^\text{max} = \frac{\text{area budget}}{M \cdot \max\{ l_{ind},l_{LS} \}}.
\end{equation}
where $\max\{ l_{ind},l_{LS} \}$ is the effective row length for each VR. 
To reduce lateral conduction losses, all VR components are vertically aligned within a single cell footprint of the $m \times n$ array, given by ${\max\{ l_{ind},l_{LS} \} \cdot w_{LS}^\text{max}}$.

When the optimal switch width, $w_{LS}^\text{opt}$ (see~(\ref{eq:wopt-lin})), exceeds the maximum allowable width, $w_{LS}^\text{max}$, the switch must be resized to fit within a single cell footprint. This enforces a tradeoff between power efficiency and footprint utilization, as undersized switches increase conduction loss but allow denser integration and greater DVPD scalability. Importantly, a geometric upper bound on switch sizing is established by~(\ref{eq:W_max}), which directly translates into a limit on the achievable current density per layer. Similar to the modular unit-inductor concept, vertical stacking of smaller parallel-connected switches can be employed to extend scalability within multi-layer packaging constraints. Furthermore, smaller components (e.g., high-side switches, capacitors) can typically be embedded within the footprint of the largest switch on adjacent layers, improving per-layer area utilization and supporting DVPD co-design efficiency.
%
\begin{comment}
\begin{equation}\label{eq:F_adj}
%\resizebox{0.8\columnwidth}{!}{$
  f_{sw}^\text{eff}=\max
  \begin{cases}
    f_{sw}^\text{min},\\
    f_{sw}^\text{adj} = \left(\frac{I_{ds}}{w_{sw}^\text{max} \cdot V_{ds}}\right)^2 {\frac{K}{\frac{C_{oss0}}{2}+C_{gs0}\left(\frac{V_{gs}}{V_{ds}}\right)^2}}.
  \end{cases}
%  $}
\end{equation}
%
\end{comment}

%Accordingly, the effective width ($w_{sw}^\text{eff}$) for each power switch in the system is determined based on $f_{sw}^\text{eff}$ and (\ref{eq:Wopt_HS}) for HS switches and (\ref{eq:Wopt_LS}) for LS switches. A floorplan for the switches and inductors layers is exemplified in Fig. \ref{fig:VR sizing}, illustrating the proposed design methodology.
%Ultimately, the optimal sizing frequency under area constraints ($f_{siz}$), is the maximum of $f_{sw}^{min}$ and $f_{adj}$, $f_{siz}=\max\{f_{sw}^{min}, f_{adj}\}$. 

The sizing factor, defined as \({w_{LS}^\text{opt}}\)/\(w_{LS}^\text{max}\), is shown in Fig.~\ref{fig:Sizing factor} as a function of the input voltage, serving as an indicator of whether switches can be optimally sized for loss minimization or are constrained by footprint area. Importantly, the sizing factor is independent of the number of distributed VRs, $M$, as long as the number of rows $m$ in the floorplanned $m \times n$ array remains unchanged. This invariance arises because \(M\), affects both \(w_{LS}^\text{opt}\) (through the reduced per-VR load current, $I_{ds}\propto \frac{1}{M}$), and \(w_{LS}^\text{max}\) (directly proportional to $\frac{1}{M}$), in the same way, and the dependencies cancel out. Consequently, the sizing factor is governed solely by the input voltage. At lower voltages ($<$\,30\,V in this example), the optimal switch width exceeds the maximum footprint-limited width (see~(\ref{eq:wopt-lin})), forcing a reduction in effective width and incurring higher conduction loss in exchange for higher current density and scalability. At higher voltages ($>$\,30\,V), the optimal width fits within the footprint, enabling switches to be sized for efficiency without sacrificing density.
\begin{figure}[b!]
    \centering
    \includegraphics[width=\columnwidth]{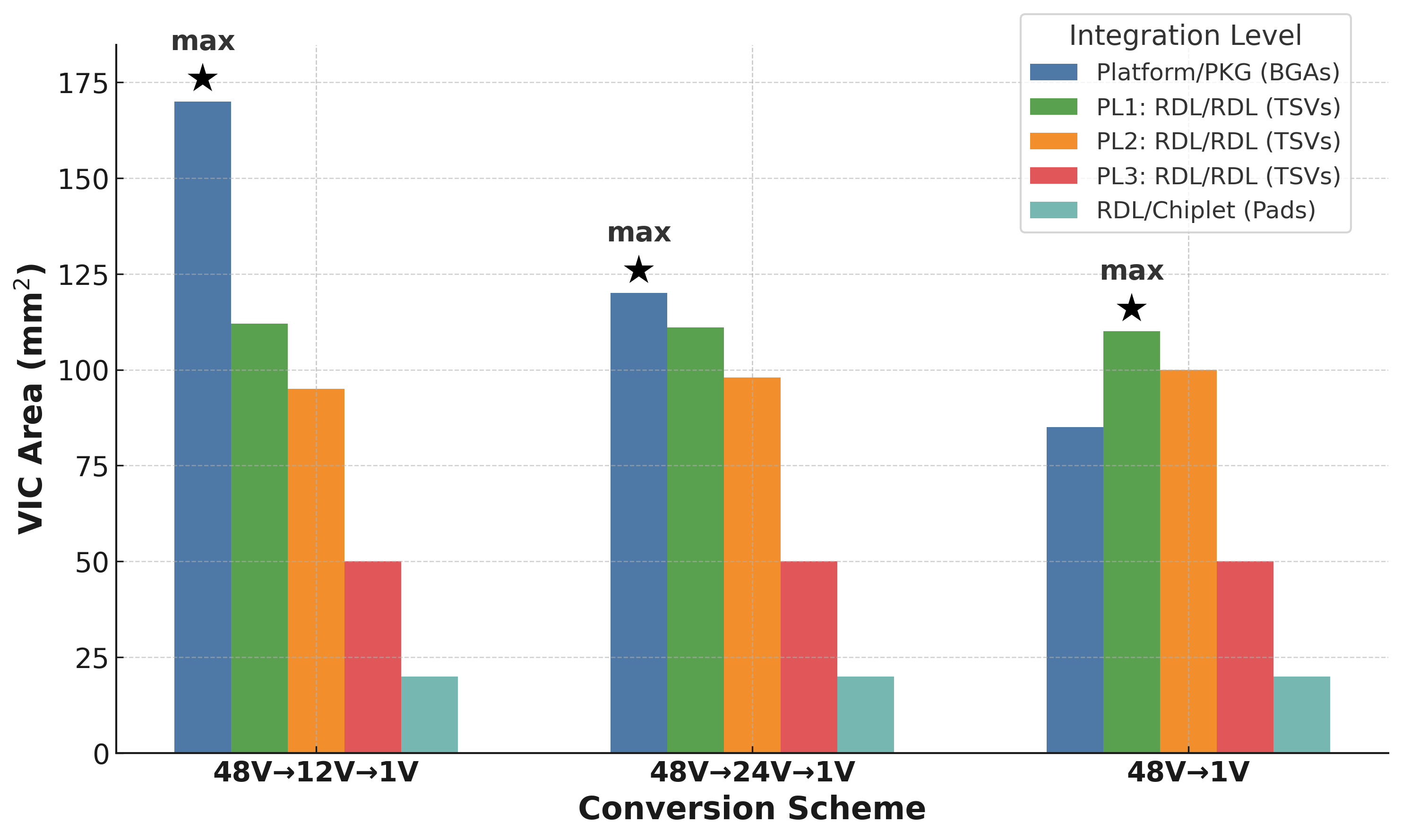}
    \caption{Per-layer VIC footprint across three voltage conversion schemes in a 1\,kW DVPD system. 
    The bottleneck layer is configuration-dependent: for direct 48$\to$1\,V conversion the coarse-pitch BGAs dominate, 
    while staged 48$\to$12$\to$1\,V operation shifts the bottleneck to TSV-based layers due to increased intermediate current.}
    \label{fig:VIC_area}
\end{figure}

Thus, the \texttt{crossover input voltage marks the boundary where footprint constraints begin to limit switch optimization}. While 30 V is specific to the 1\,kW example in Fig.~\ref{fig:Sizing factor}, the crossover point is design-dependent and can be determined using the proposed framework for any performance target and device technology.

The proposed DVPD floorplanning further reduces power dissipated in PDN between the VR components. Specifically, the effective width (and thus, per-length resistance) of the internal VR interconnect is roughly determined by the length of the largest VR component, $\max\{l_{ind},l_{sw}\}$ and is independent of M. Alternatively, the effective routing length between VR components is defined by $w_{LS}^{opt}\propto1/M$, as illustrated in Fig.\ref{fig:Routing_length_scaling}. 

\texttt{Combined with the per-VR DVPD current reduction, the proposed approach achieves $M(R_{PDN}/M) (I/M)^2 \propto 1/M^2$ reduction in the internal VR routing losses}. 

%\hl{Note that this intra-VR routing loss reduction is independent of the load system aspect ratio, and is governed solely by intra-VR routing geometry. The $1/M^2$ scaling holds under the following conditions: (1)~the $M$ VRs form a regular array across the system footprint, (2)~the effective routing width is set by $\max\{l_{ind}, l_{sw}\}$, independent of $M$, and (3)~each VR is rated for $1/M^{th}$ of the maximum load current.}

Note that this intra-VR routing loss reduction is independent of the overall load-system footprint geometry and is governed by the local intra-VR routing geometry. The derivation assumes that (1) the distributed VR cells are locally similar as $M$ increases, (2) the effective routing width is determined by $\max\{l_{ind},l_{sw}\}$ and remains approximately independent of $M$, (3) each VR is designed to deliver approximately $1/M$ of the maximum load current, and (4) intra-VR routing is not constrained by local blockages, keep-out regions, or package design-rule limitations that would force routing detours or reduce available routing bandwidth.

\subsection{Interconnect}\label{subsec:VIC}

\begin{table}[t]
    \caption{End-to-End Efficiency Boundary and Loss Accounting}
    \label{tab:efficiency_boundary}
    \centering
    \renewcommand{\arraystretch}{1.5}
    \setlength{\tabcolsep}{2.6pt}
    \begin{tabular}{p{1.7cm}|p{1.6cm}|p{1.5cm}|p{1.5cm}|p{1.5cm}}
        \toprule
        \textbf{Power Segment} &
        \textbf{Baseline} &
        \textbf{A1} &
        \textbf{A2} &
        \textbf{A3} \\
        \midrule
        System-level boundary &
        \multicolumn{4}{c}{48-V Board-Level Input $\rightarrow$ POL} \\
        \midrule
        Conversion scheme &
        48V/12V/1V & 48V/1V &
        48V/24V/1V & 48V/12V/1V \\
        
        Platform-level VR &
        48V/12V XFMR$^1$, 12V/1V buck &
        N/A &
        48V/24V XFMR$^1$ &
        48V/12V XFMR$^1$ \\
        
        Platform-level PDN &
        Included &
        Included &
        Included &
        Included \\
         
        Embedded/ \hspace{15pt} package VR &
        N/A &
        Hybrid$^2$ &
        Hybrid$^2$ &
        Hybrid$^2$ \\
         
        Package-level \hspace{5pt} PDN$^3$: BGAs, TSVs, RDLs,  chiplet pads) &
        Included &
        Included &
        Included &
        Included \\
         
        % \midrule
        % Estimated \hspace{15pt} efficiency &
        % 66\% &
        % 84\% &
        % 86\% &
        % 87.6\% \\
        \bottomrule
        \multicolumn{5}{l}{$^1$A 98\%-efficient transformer-based converter is assumed for illustration;\vspace{-5pt}}\\
        \multicolumn{5}{l}{other high-efficiency front-end converters can be used and are supported.\vspace{-2pt}}\\
        \multicolumn{5}{l}{$^2$Hybrid VRs are assumed for illustration; resonant converters are also \vspace{-5pt}}\\
        \multicolumn{5}{l}{promising; other VRs are also supported.\vspace{-2pt}}\\
        \multicolumn{5}{l}{$^3$See Table~\ref{tab:vic_char}.}
    \end{tabular}
\end{table}

Efficient DVPD requires co-optimization of vertical and lateral interconnects to satisfy current demands without exceeding the available in-package area. Vertical interconnects (VICs), implemented through BGAs, TSVs, or fine-pitch pads depending on the integration level, must be sized to carry the required current while ensuring mechanical robustness and electrical reliability. Different nodes within a VR cell carry different currents; the number of parallel VICs per connection is sized based on the connection current and per-via current limit, with current assumed to be uniformly distributed across the allocated VICs. A critical observation is that the performance-limiting layer is not determined solely by the layer carrying the largest current. Advanced packaging technologies enable fine-pitch vertical interconnects (down to $<$10\,\textmu m pitch) supported by interfaces such as SuperChips ~\cite{jangam2021silicon} and BoW ~\cite{farjadrad2019bunch}), allowing dense high-current delivery in upper layers. By contrast, bottom layers such as BGA interfaces rely on coarse-pitch solder balls, consuming significantly more area per via despite carrying lower current. Thus, the bottleneck layer is jointly determined by current magnitude, conversion ratio ($V_{out}/V_{in}$), and interconnect technology.

\setlength{\tabcolsep}{0.45em} % for the horizontal padding
{\renewcommand{\arraystretch}{1.0}
\begin{table}[b!]
  \caption{Performance of Embedded Inductors.}
  \begin{center}
  \vspace{-5pt}
  \setlength\tabcolsep{2.6pt}
  \resizebox{\columnwidth}{!}{%
  \begin{tabular}{l|l|c|c|c|c|c|c}
      \hline
       Type & Reference & Core & $L$ & $L_\square$ & $I_{\text{sat}}$ & $I_\square$ &  $R_{\text{ESR}}$  \\
      %\hline
      & & $\mu_r$ & ({nH}) & ({nH}/mm$^2$) & (A) & ({A}/mm$^2$) &  (m$\Omega$)\\
      \hline
      PCB & APEC'14\cite{burton2014fivr} & 1 & 1.2 & 0.6 & 8 & 4 & 7 \\
      Coax & ECTC'21\cite{bharath2021integrated} & 8.5 & 2.5 & 6.3 & 8 & 20 & 12 \\
      Toroid & ECTC'21\cite{murali2022fabrication} & 30 & 420 & 48 & 2.5 & 0.3 & 89\\
      Spiral & ECTC'25\cite{rasheedi2025high}\textsuperscript{a} & 
      %4.7 & 78 & 11.1 & 24 & 3.4 & 11 \\
      15 & 107 & 15.2 & 70 & 9.9 & 11 \\
      Solenoid & T-ED'24 \cite{pan2024high} & - & 340 & 56.7 & 2.2 & 0.4 & 308 \\
      Solenoid & T-ED'21 \cite{he2021chip}& - & 6.5 & 14.4 & 1.6 & 3.6 & 22 \\
      TSV & TPEL'26 \cite{ding2026through}\textsuperscript{b} & 600 & 20 & 54 & 0.5 & 1.35 & 36 \\
      Spiral & TCPMT'25 \cite{avula2024design} & 180 & 330 & 36.7 & 3.5 & 0.39 & 85 \\
      Array & TCPMT'21 \cite{barros2021embedded} & 65 & 170 & 33.6 & 1.9 & 0.37 & 15 \\
      
      Target & - & - & 250 & 77 & 5 & 1.5 & 12 \\
       \hline
    \end{tabular}}

    \begin{tablenotes}
        \item \hspace{-5pt}\textsuperscript{a}Re-simulated with TY-M5 material.
        \item \hspace{-5pt}\textsuperscript{b}Ten TSVs connected in series.
    \end{tablenotes}
    \label{tab:Inductors}
  \end{center}
  %\begin{tablenotes}
   % \vspace{-5pt}
   % \item $^a$Single spiral 
    %\tabto{9em} $^b$Two parallel spiral 
    %\tabto{18em} $^c$Progressively widened spiral
    %\vspace{-8pt}
   % \item $^d$Single composite \tabto{9em} $^e$Dual composite
  %\end{tablenotes}
%  \vspace{-10 pt}
\end{table}

The scaling with input voltage can be approximated analytically. Since RMS current directly governs electromigration and power loss in the interconnects and defines the current-carrying requirement and long-term reliability of each vertical link, it provides the appropriate basis for estimating VIC demand and footprint. For buck-derived (including hybrid) converters in continuous conduction mode, the input RMS current can be expressed as
\begin{equation}
    I_{in,rms} \propto I_{out} \sqrt{\tfrac{V_{out}}{V_{in}}},
\end{equation}
valid when the input current waveform is nearly a square pulse with duty cycle proportional to $V_{out}/V_{in}$. For example, reducing the VR input voltage by a factor of four, from 48 V to 12 V (while maintaining a 1 V output), increases the RMS input current by a factor of two, even though the average input current rises by four. Since the required VIC count and footprint area scale with $I_{in,rms}$, the overall interconnect area likewise doubles. This square-root dependence provides a fast rule of thumb for DVPD planning: lowering the VR input voltage improves conversion efficiency but simultaneously increases interconnect demand, reducing the in-package real estate available for VR embedding.  

The per-layer VIC areas for representative conversion scenarios are illustrated in Fig.~\ref{fig:VIC_area} (additional details are provided in Section~\ref{sec:Results}). The results show that the layer dictating system scalability shifts depending on the conversion ratio: in direct 48\,V$\to$1\,V conversion, the TSV-based layers %coarse-pitch BGAs%
dominate the footprint, whereas in staged 48\,V$\to$12\,V$\to$1\,V operation, the higher intermediate VR input current drives coarse-pitch BGAs in bottom layer %TSV-based layers% 
toward the area limit. This highlights a fundamental DVPD tradeoff: \texttt{interconnect bottlenecks are configuration-dependent} and must be evaluated through joint consideration of current scaling and packaging technology.  

\begin{figure}[t!]
    \centering
    \includegraphics[width=\columnwidth]{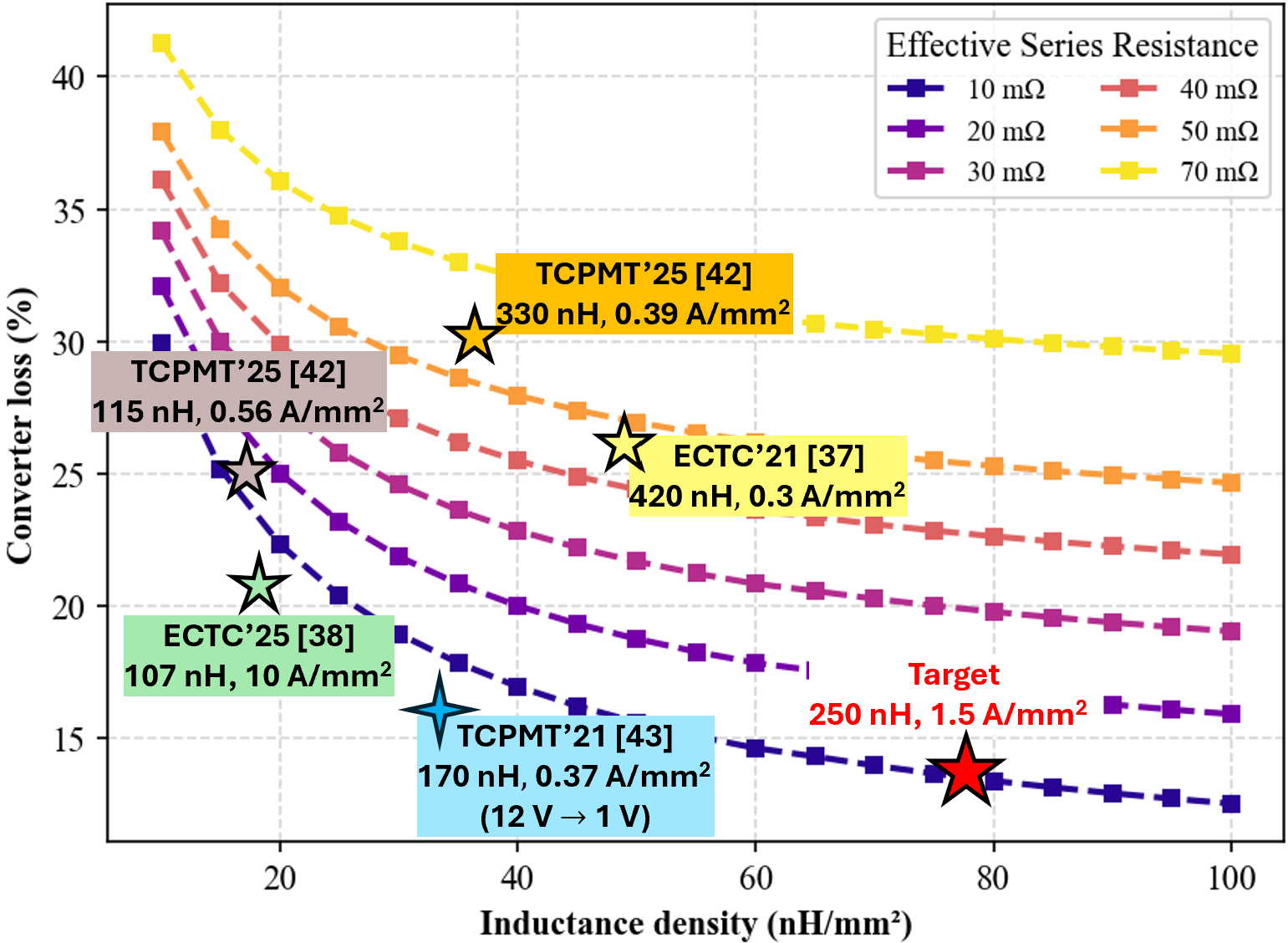}
    \caption{Sensitivity of VR power loss to embedded inductor characteristics, showing VR loss as a function of inductor inductance density and DC resistance as extracted from the proposed DVPD framework. Representative recently demonstrated embedded inductors from Table I are overlaid for comparison. The results highlight the tradeoffs among inductance density, resistance, and current density, and identify the target operating region required to achieve high-efficiency DVPD operation.}
    \label{fig:sensitivity}

\end{figure}

Lateral redistribution layers (RDLs) complement vertical interconnects by routing power and ground both within and across VRs. Although the specific trace dimensions depend on process technology, the essential observation is that \texttt{allocating separate RDL layers to intra-VR and inter-VR distribution} enables balanced current allocation and preserves signal integrity. Taken together, the vertical interconnect capacity and the lateral routing density define the achievable current density and integration limits of the DVPD architecture.

\section{DVPD Design Space Exploration and Framework Validation}\label{sec:Results}

To evaluate the efficiency and density limits achievable with practical device and packaging technologies, 
the proposed framework is applied to a 1\,kW, 2\,A/mm\textsuperscript{2} DVPD system employing DSCH regulators. A representative DVPD VR-cell implementation used in the following case studies is illustrated in Fig.~\ref{fig:VR-cell}. %\hl{The figure summarizes the assumed power-layer organization, embedded components, and vertical interconnect hierarchy utilized to instantiate the framework.}

For consistency, all efficiency values reported in this section are evaluated using a unified end-to-end power delivery boundary extending from platform-level input power to delivered on-chip load power. Accordingly, all reported efficiencies include front-end conversion losses, VR conversion losses, package/interconnect distribution losses, and VR-to-POL routing losses. The corresponding loss accounting for the evaluated architectures is summarized in Table~\ref{tab:efficiency_boundary}.

Representative state-of-the-art inductors across different topologies are compared in Table~\ref{tab:Inductors}. While certain structures can achieve high current density, this is often accompanied by reduced inductance density, increased area, or higher resistance, reflecting a fundamental tradeoff among inductance density, current capability, and loss in embedded magnetic structures.

Based on representative recently demonstrated embedded inductors (see Table~\ref{tab:Inductors}), a target unit inductor specification of 77 nH/mm² inductance density, 5 A peak current, and 11\,m$\Omega$ DC resistance is adopted in this study to evaluate the achievable DVPD design space. 
%These values do not correspond to a single experimentally demonstrated device, but rather represent a near-term design target extrapolated from recent trends in embedded magnetic inductor technologies.

%\hl{To further assess sensitivity to embedded inductor technology}, Fig.~\ref{fig:sensitivity} \hl{presents VR loss as a function of inductance density and effective series resistance together with representative recently demonstrated embedded inductors. The results highlight the dependence of achievable DVPD efficiency on embedded inductor characteristics and demonstrate how the proposed framework can quantitatively determine the passive-device characteristics required to achieve target system-level efficiency and density goals.}
%Note that the proposed DVPD framework is agnostic to inductor type and can evaluate system-level performance for any specified inductor parameters. 

Fig.~\ref{fig:sensitivity} presents VR loss as a function of inductance density and effective series resistance together with representative embedded inductors. The results highlight the dependence of achievable DVPD efficiency on inductor characteristics and identify the passive-device requirements needed to achieve target system-level efficiency and density goals. The framework is agnostic to inductor type and can evaluate performance for arbitrary inductor parameters.

\begin{table}[b!]
\centering
\caption{Vertical Interconnect Characteristics and Optimized VIC Allocation Across Integration Levels.}
\label{tab:vic_char}
\setlength{\tabcolsep}{2.4pt}
\renewcommand{\arraystretch}{1.15}
\footnotesize

\begin{tabular}{c|l c c c c c}
\toprule
%& \makecell[lb]{\textbf{Integration}\\\textbf{level}}
& & \makecell[b]{\textbf{Platform$^a$}\\\textbf{/PKG}}
& \makecell[b]{\textbf{PL$_1$}\\\textbf{RDL/}\\\textbf{RDL}}
& \makecell[b]{\textbf{PL$_2$}\\\textbf{RDL/}\\\textbf{RDL}}
& \makecell[b]{\textbf{PL$_3$}\\\textbf{RDL/}\\\textbf{RDL}}
& \makecell[b]{\textbf{RDL/}\\\textbf{Chiplet}} \\
%\midrule
\cline{2-7}

% Top-aligned, but with small top margin
\multirow[t]{7}{*}{\makecell{\\\\\\\\\\\textbf{Technology}\\\textbf{parameters}}}
& Type        & BGAs & TSVs & TSVs & TSVs & Pads \\
& Material    & solder$^b$ & Cu$^b$ & Cu$^b$ & Cu$^b$ & Cu$^b$ \\
& $d_c$ (\textmu m) & 400 & 30 & 30 & 30 & 10 \\
& $h_c$ (\textmu m) & 300 & 300 & 300 & 300 & 10 \\
& $p_c$ (\textmu m) & 800 & 60 & 60 & 60 & 20 \\
& $A_c$ (\textmu m$^2$) & 125{,}663 & 707 & 707 & 707 & 100 \\
& $R_c$ (m$\Omega$) & 0.3 & 7.1 & 7.1 & 7.1 & 1.7 \\
\midrule

\textbf{48V/1V}
& \# VICs       & 132 & 30{,}202 & 27{,}845 & 14{,}148 & 200{,}000 \\
& Area (mm$^2$)  & 85  & 110      & 100      & 50       & 20 \\
\midrule

\textbf{48V/24V/1V}
& \# VICs       & 186 & 30{,}715 & 27{,}380 & 14{,}148 & 200{,}000 \\
& Area (mm$^2$)  & 120 & 111      & 98       & 50       & 20 \\
\midrule

\textbf{48V/12V/1V}
& \# VICs       & 266 & 31{,}115 & 26{,}399 & 14{,}148 & 200{,}000 \\
& Area (mm$^2$)  & 170 & 112      & 95       & 50       & 20 \\
\bottomrule
\end{tabular}

\vspace{4pt}

\begin{minipage}{\columnwidth}
\footnotesize\raggedright
$^a$ PCB, wafer, panel or any other system integration platform.\\
$^b$ Solder resistivity: 130 n$\Omega\cdot$m, Cu resistivity: 17 n$\Omega\cdot$m.
\end{minipage}
%\vspace{-10pt}

\end{table}

Flying capacitors with voltage stress levels of 8--32\,V and a capacitance density of 1--2\,\textmu F/mm\textsuperscript{2} are employed \cite{khorasani2025embedded}. Each VR cell incorporates two 3.2\,\textmu F flying capacitors sized to satisify the 2\% ripple constraint and embedded within the middle layer ($\text{PL}_2$ in Fig. \ref{fig:DVPD}) as part of the local VR-cell footprint.
VIC characteristics are selected based on representative fabrication constraints reported in \cite{safari2023robust},\cite{khorasani2025embedded}, including aspect ratio, pitch, and material properties, and are listed in Table~\ref{tab:vic_char}. These parameters define a feasible design point for evaluation; however, the proposed framework is not limited to these values and can incorporate alternative interconnect technologies and specifications.
%VIC characteristics are determined based on \cite{safari2023robust}\cite{khorasani2025embedded} and are summarized in Table~\ref{tab:vic_char}. 
Representative interconnect dimensions and the corresponding vertical interconnect hierarchy are illustrated in Fig.~\ref{fig:VR-cell}. The fine-pitch 10 \textmu m interfaces correspond to chiplet-pad connections, whereas the substrate TSVs employ substantially larger dimensions consistent with current glass and silicon substrate technologies. %The fine-pitch 10 \textmu m interfaces correspond to chiplet-pad connections, whereas the substrate TSVs employ substantially larger dimensions consistent with current glass and silicon substrate technologies.

\begin{figure}[t!]
    \centering
\includegraphics[width=0.7\columnwidth]{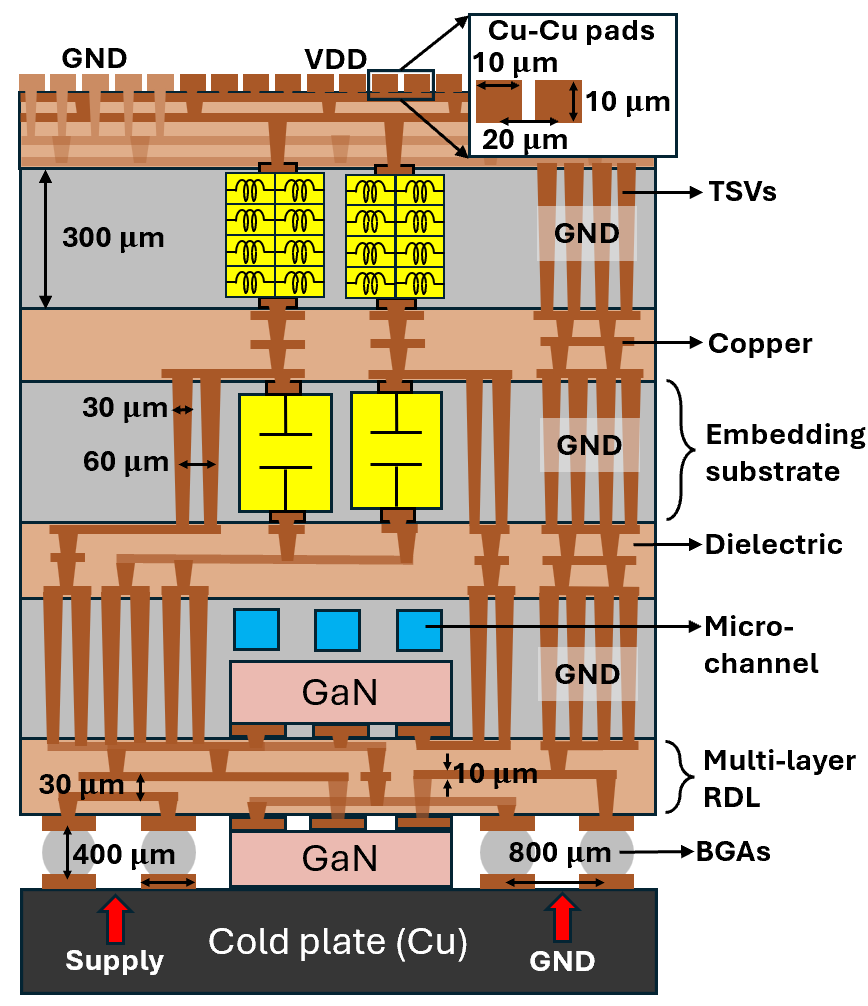}
    \caption{Representative DVPD VR-cell implementation used in the case studies, illustrating stacked power-delivery layers, embedded active and passive components, vertical interconnect hierarchy, and representative dimensions for BGAs, TSVs, and chiplet-interface pads.}
    \label{fig:VR-cell}
     \vspace{-10pt}
\end{figure}

GaN devices with \(V_{gs} = 5\,\text{V}\),
\(\mu C_\text{AlGaN} = 67.4\,\text{A/V}^2\),
\(l_\text{sw} = 1\,\text{mm}\),
\(C_\text{iss0}~=~220\,\text{pF/mm}^2\),
\(C_\text{oss0} = 150\,\text{pF/mm}^2\),
and \(I_\text{g} = 2.4\,\text{A}\) are considered. 
The parameters are extracted from SPICE \(I_{ds}\)–\(V_{gs}\) characteristics
and EPC GaN device datasheets~\cite{epc2023gan}
and are used to construct the corresponding switch-loss models within the framework.
Once the target performance, technology specifications, and VR topology are provided as inputs, the framework automatically performs the corresponding DVPD design-space exploration and optimization, generating detailed per-layer floorplans 
together with comprehensive electrical and physical performance report. 
These reports include spatial and component-level power-loss distributions, steady-state and transient voltage drops, 
and density characteristics used to identify architectural bottlenecks and quantify overall system-level tradeoffs.  

Note that the results presented in this section are not intended to represent a single globally optimal DVPD system. Rather, they demonstrate the capability of the proposed framework to systematically evaluate DVPD design tradeoffs under a representative set of technology and architectural assumptions.

\begin{figure}[b!]  \centerline{\includegraphics[width=0.45\textwidth]{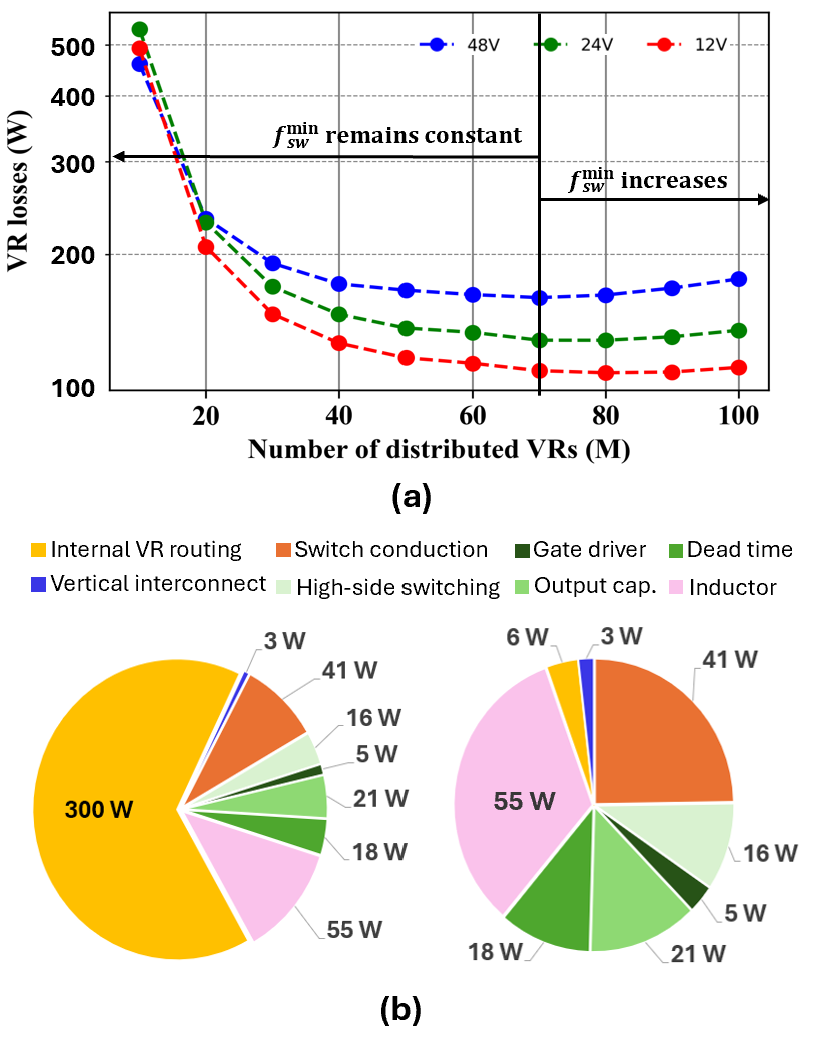}}
  \caption{ Total conversion loss and loss distribution in a DVPD system delivering 1 kW at 1 V to a 500 mm\textsuperscript{2} load system.
(a) Total conversion loss as a function of the number of distributed VRs.
(b) Loss distribution for 10-VR (left) and 70-VR (right) configurations.
}
\label{fig:Conversion loss}

\end{figure}

\subsection{Conversion Loss}

The total VR losses---combined switching and conduction losses of all stacked components (switches, inductors, capacitors) 
and their interconnections---are shown in Fig.~\ref{fig:Conversion loss}. Increasing the number of VRs reduces the per-VR current linearly, while the lateral routing distance between VR components also decreases as the VR footprint shrinks (see Section~\ref{subsec:Sizing}). Together, these effects reduce intra-VR routing losses, which scale quadratically, thereby improving the overall power conversion efficiency. While the demonstrated power-loss savings are lower than the predicted theoretical value ($M^2$) due to additional loss components associated with inductors and power-switches that do not scale with $M$, significant power-loss reduction is still observed with DVPD. However, once the per-phase current falls below the saturation rating of a unit inductor, higher switching frequencies are required to satisfy ripple constraints (see Fig.~\ref{fig:Minimum switching frequency}), which increases switching losses. This tradeoff establishes an optimal VR count---70 in the DSCH-based case study---that maximizes system efficiency.

Rather than being prescribed analytically, the preferred VR count is obtained by numerically minimizing the total modeled VR-system loss over feasible integer values of $M$. For each candidate $M$, the framework optimizes the design parameters described in Section \ref{sec:Methodology}, including switching frequency, modular inductor allocation, switch sizing, floorplanning, and interconnect allocation, and evaluates the resulting combined switching, conduction, and routing losses~\cite{krishnakumar2024vertical,jauregui2011power} with the optimized design parameters subject to footprint and ripple constraints. Thus, the optimal value of $M=70$ in this case study is a design-point-specific output of the framework rather than a general result.

To further assess sensitivity to power-switch technology, the framework is evaluated across switch channel lengths ($l_\text{sw}$) ranging from 0.5 mm to 1.8 mm for the DVPD system with 48 V input voltage, 70 distributed VRs, and a nominal switch channel length of $l_\text{sw}$ = 1.0 mm. Since $l_\text{sw}$ scales both device on-resistance and capacitance in the adopted switch model, this sweep captures the combined sensitivity of conduction and switching losses. Conversion loss increases from 10.87\% at $l_\text{sw}$ = 0.5 mm to 13.88\% at the nominal $l_\text{sw}$ = 1.0 mm and 18.68\% at $l_\text{sw}$ = 1.8 mm (Table~\ref{tab:channel}).

\begin{table}[t!]
\centering
\fontsize{7}{9}\selectfont
\caption{Sensitivity to Power-Switch Technology}
\label{tab:channel}
\begin{tabular}{ccc}
\hline
\textbf{Channel length} & \textbf{Conversion loss} & \textbf{Efficiency} \\
\textbf{(mm)} & \textbf{(\%)} & \textbf{(\%)} \\
\hline
0.5 & 10.87 & 87 \\
1.0 & 13.88 & 84 \\
1.8 & 18.68 & 80 \\
\hline
\end{tabular}
\vspace{-5pt}
\end{table}

\begin{table}[t!]
\centering
\caption{Sensitivity to Interconnect Technology}
\label{tab:interconnect}
\resizebox{\columnwidth}{!}{%
\begin{tabular}{lllccc}
\hline
\textbf{Design} & \textbf{Interconnect} & \textbf{Bottleneck} & \textbf{VIC area} & \textbf{VIC loss} & \textbf{Efficiency} \\
 \textbf{case}& \textbf{assumption} & \textbf{layer} & \textbf{(mm\textsuperscript{2})} & \textbf{(W)} & \textbf{(\%)} \\
\hline
Nominal & Baseline & $\text{PL}_1$: RDL/RDL & 110 & 3.71 & 84.46 \\
TSV-R $\uparrow$ & $R_\text{TSV} = 1.5\times$ & $\text{PL}_1$: RDL/RDL & 110 & 4.12 & 84.28 \\
TSV pitch $\uparrow$ & $\rho_\text{TSV} = 1.5\times$ & $\text{PL}_1$: RDL/RDL & 245 & 3.71 & 84.64 \\
BGA pitch $\uparrow$ & $\rho_\text{BGA} = 1.5\times$ & Platform/PKG & 191 & 3.71 & 84.46 \\
\hline
\end{tabular}%
}
\end{table}

To assess sensitivity to interconnect technology, the framework is further evaluated by varying TSV resistance, TSV pitch, and BGA pitch across representative ranges. The resulting bottleneck layer, interconnect loss, and end-to-end efficiency are summarized in Table~\ref{tab:interconnect}. The results indicate that the primary conclusions remain robust across the evaluated range of interconnect technology assumptions.

\subsection{VR Placement and Spatial Effects}

The geometry of the VR cell also impacts performance. 
Among the possible $m \times n$ array configurations within the system area, \texttt{balanced cells} (aspect ratio $\approx 1$) are preferred, 
as they minimize elongated RDL routes and reduce distribution loss and minimize voltage drops. %\hl{For a non-square load systems, the $m \times n$ array aspect ratio should be matched to that of the load system footprint, ensuring that individual VR cells remain approximately square and VR-to-POL routing distances are uniformly distributed across the system.}
For non-square load systems, the $m \times n=M$ array aspect ratio should be matched as closely as practical to that of the load-system footprint, ensuring that VR-to-POL routing distances remain more uniformly distributed across the system. The footprint of each cell is determined by the largest components (switches or inductors). 
Accordingly, these components should be carefully selected or designed to enable balanced cells, 
ensuring both efficient routing and uniform distribution. 
The optimized floorplan for 70 VRs is shown in Fig.~\ref{fig:Optimized VR placement}.

Finally, spatial variation across the load system is captured by partitioning the chiplet into multiple POL regions. 
While a fully detailed model would require treating each transistor as an individual load, in practice the system can be 
partitioned into a modest number of regions with high accuracy \cite{krishnakumar2024system}. In this work, the load is divided into 100 POL segments, 
as illustrated in Fig.~\ref{fig:Optimized VR placement}. Current is primarily supplied by nearby VRs, which creates a 
dual behavior across the system: VRs near the center supply multiple surrounding POLs and exhibit higher conversion losses, 
while VRs near the perimeter carry less current and show reduced loss \cite{kose2010fast}. Conversely, central POLs benefit from multiple 
parallel current paths, whereas corner POLs have fewer nearby VRs, leading to nearly $2\times$ higher routing losses. 

\begin{figure}[b!]

\centerline{\includegraphics[width=0.5\textwidth]{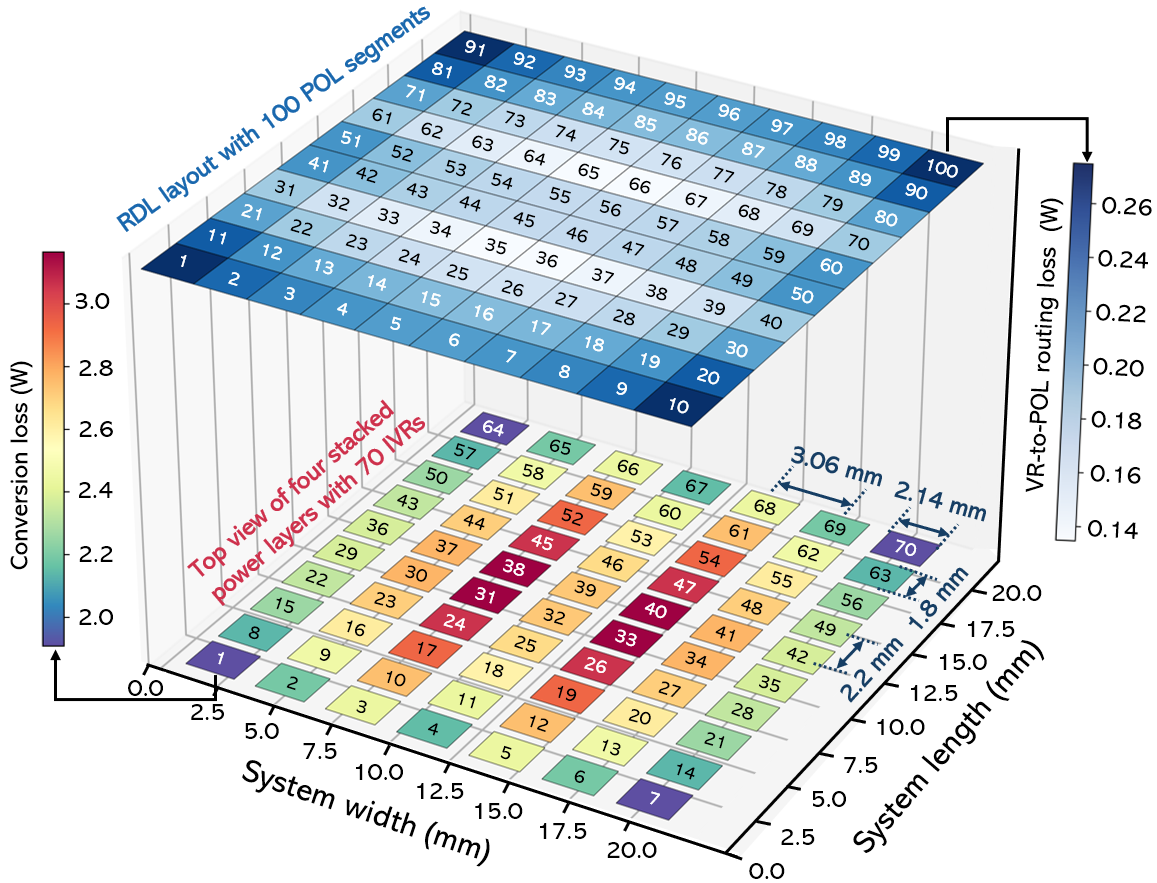}}
\caption{Optimized floorplan designed with the automated DVPD framework to deliver 1\,kW at 1\,V, to a 500\,mm\textsuperscript{2} load system, illustrating conversion and routing loss distribution.}
\label{fig:Optimized VR placement}

\end{figure}

The spatial voltage drop across the system for VPD with a single VR and DVPD with 70 distributed VRs is illustrated in Fig.~\ref{fig:Voltage_drop}a and Fig.~\ref{fig:Voltage_drop}b, respectively. %DVPD reduces the voltage drop by approximately 1.5$\times$ with \textit{IR} drops below 2.7\% and \textit{L}\,d\textit{i}/d\textit{t} drops below 9\%. Strategically placed edge decoupling capacitors will further reduce the worst-case voltage drop to below 9\%.
DVPD reduces the voltage drop by approximately 1.5× with IR drops below 2.7\% and $L \frac{di}{dt}$ drops below 9\%. Strategically placed edge decoupling capacitors can further reduce the worst-case transient voltage drop. As a representative feasibility estimate, for the evaluated 1 kW, 1 V system, supporting 25–33\% of the full-load transient current using deep-trench capacitors with reported capacitance densities of at least 100 fF/µm\textsuperscript{2}~\cite{kannan2020deep} requires approximately 28–37 mm\textsuperscript{2} (5.6–7.4\% of the 500 mm\textsuperscript{2} load footprint). Since the output-capacitor layer remains substantially underutilized in the evaluated DVPD configurations (see Fig. \ref{fig:DVPD}), this capacitance can be accommodated with limited impact on the overall footprint. Assuming the transient voltage drop scales approximately with the unsupported transient current, supporting 25–33\% of the full-load transient current is estimated to reduce the worst-case transient voltage drop from 9\% to ~6–7\%.

\subsection{Power Loss Distribution}

System-wide power loss distribution for the traditional and DVPD architectures, evaluated using the unified end-to-end efficiency boundary defined in Table~\ref{tab:efficiency_boundary}, is shown in Fig.~\ref{fig:comparison}. The traditional reference architecture employs a two-stage 48\,V$\to$12\,V$\to$1\,V conversion chain, consisting of a transformer-based front-end converter and a multi-phase motherboard buck converter with an assumed total 90\% VR conversion efficiency. Once the VR-to-POL lateral routing losses are included within the unified end-to-end efficiency boundary, the overall system efficiency falls below 70\% (see Fig.~\ref{fig:comparison}). %\hl{consistent with reported measurements for representative power delivery solutions (e.g., NVIDIA V100) in }\cite{krishnakumar2023vertical,khairy2020}.

%\sk{A single-VR VPD architecture posses several practical limitations. To sustain high current delivery, excessively wide power switches violating area constraints, and air-core inductors preventing magnetic saturation are required. However, such switches, optimized for conduction loss, and inductors with low inductance density fail to meet the high-current-density requirements of modern HPC systems. Consequently, multiple parallel-connected off-the-shelf power switches and magnetic-core unit inductors must be employed, resulting in conduction and switching losses each exceeding 20\% of the total power delivered and yielding a system-wide efficiency of approximately 62\%. Hence, in this work, eight large optimized DSCH VRs (approximately 125 A each) are employed in the VPD baseline to realize a practically feasible architecture from a power-efficiency standpoint.}
%The VPD baseline integrates eight large DSCH VRs (approximately 125\,A each) flip-chipped beneath the load system. 
The DVPD configurations, $A_1$, $A_2$, and $A_3$, employ 70 horizontally distributed VRs with vertically stacked components, 
operating with 48\,V, 24\,V, and 12\,V inputs, respectively. For $A_2$ and $A_3$, the first-stage conversion (48\,V$\to$24\,V and 48\,V$\to$12\,V, respectively) is performed using highly efficient ($\sim$98\%) transformer-based platform converters, and the corresponding 2\% loss is included in the overall VR conversion losses.
Accordingly, all architectures are evaluated using a consistent end-to-end accounting methodology, enabling direct comparison of system-wide power delivery efficiency.

\begin{figure}[t!]

\centerline{\includegraphics[width=0.5\textwidth]{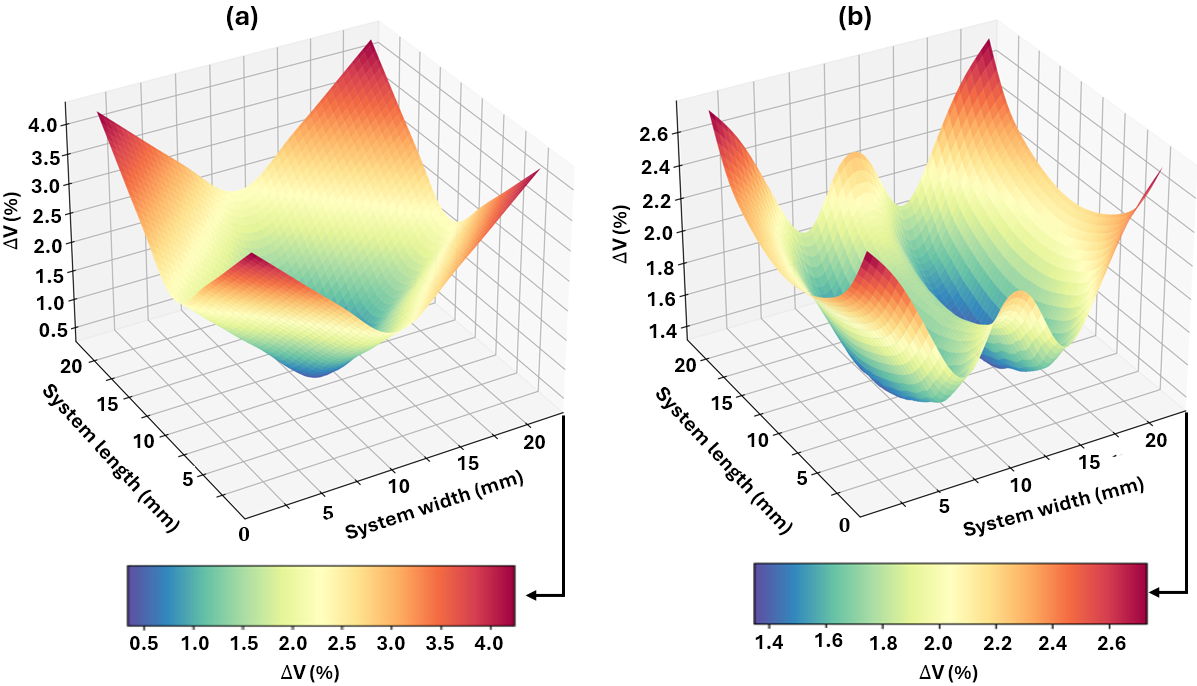}}
\caption{Spatial voltage drop across a 1-kW, 1-V load system:
(a) VPD architecture with a single VR; (b) DVPD architecture with 70 distributed VRs.}
\label{fig:Voltage_drop}
\vspace{-10pt}
\end{figure}
Among the DVPD variants, $A_2$ (24\,V$\to$1\,V) and $A_3$ (12\,V$\to$1\,V) achieve higher efficiencies of 86\% and 87.6\%, respectively. 
However, these configurations require nearly 75\% of the load-die area for VICs and routing, 
raising concerns regarding fabrication reliability and thermal stress. 
By comparison, $A_1$ (48\,V$\to$1\,V) achieves 84\% end-to-end efficiency while occupying only 54\% of the load-die area, 
representing the most balanced design in terms of efficiency, scalability, and integration feasibility. These results demonstrate that distributed and vertically stacked architectures are able to significantly reduce losses 
while maintaining practical footprint and packaging constraints.

%\hl{To further illustrate the framework's design space exploration capability, $A_1$ is evaluated with the demonstrated inductor in }\cite{rasheedi2025high}, \hl{yielding an efficiency of 75.8\%, exceeding the $<$70\% efficiency inherent in conventional architectures.}

%The proposed DVPD approach simultaneously mitigates the high VR-to-POL routing losses observed in the conventional architecture and the high conversion losses inherent to the lumped VPD architecture, suppressing the system-wide power loss to below 20\% with $A_1$ and below 15\% with $A_3$. The internal VR routing loss decreases linearly with the number of VRs, for example, an 8.75$\times$ increase in VR count with DVPD, compared to VPD, yields a proportional reduction in routing loss. In contrast, the power switch conduction loss does not scale linearly with VR count, as the reduction from lower per-VR load current is partially offset by the increase in switch on-resistance ($R_{ds} \propto 1/w_{sw}$) with higher numbers of VRs. 
%The inductor conduction loss remains unchanged in both systems because identical inductors are used and the total number of unit inductors in the array remains constant as long as the net load current is fixed.

\subsection{Thermal Feasibility}

The thermal behavior of the DVPD system is dominated by the power switches, which account for more than 50\% of the conversion loss in $A_1$ (Fig. ~\ref{fig:Conversion loss}(b)). Consequently, the spatial non-uniformity in per-VR losses shown in Fig.~\ref{fig:Optimized VR placement} results in non-uniform heating, with hotspots concentrated near the center of the DVPD footprint and within the GaN switch layers. Under isothermal conditions, the 70-VR $A_1$ configuration exhibits a total system loss of 185 W, corresponding to an end-to-end efficiency of 84.6\%. Without thermal management, the resulting junction temperature exceeds 110°C, increasing temperature-dependent conduction and switching losses to 225 W and reducing efficiency to 81.6\%. These results highlight the importance of the electrothermal co-design loop shown in Fig.~\ref{fig:em-frame}.

\begin{figure}[b!]
\centerline{\includegraphics[width=\columnwidth]{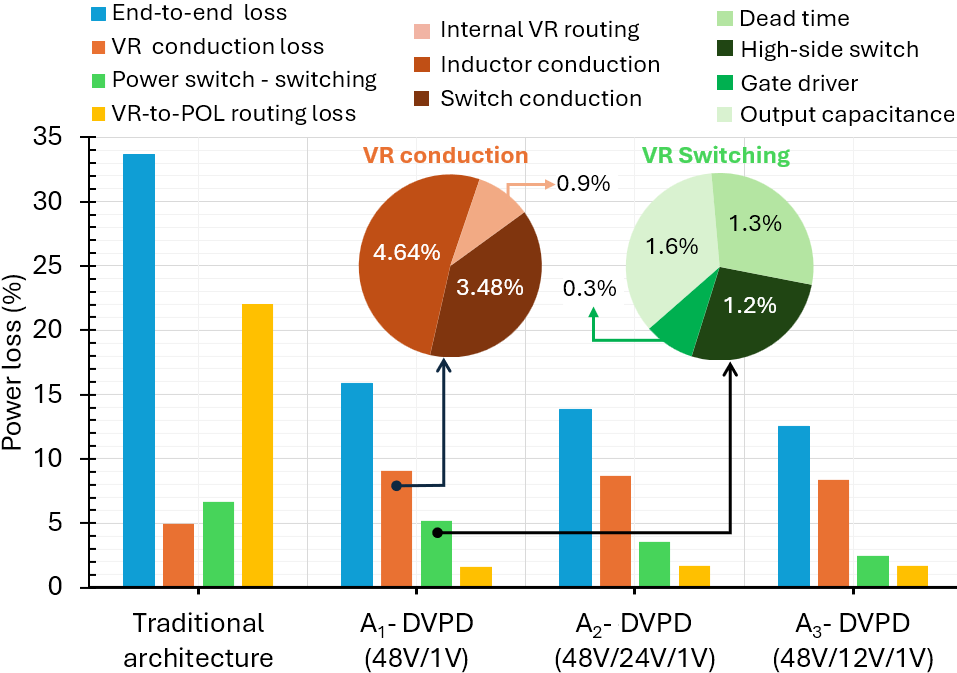}}
\caption{End-to-end power loss across traditional and DVPD architectures evaluated using the unified efficiency boundary defined in Table~\ref{tab:efficiency_boundary}. Bars show the percentage of total platform input power dissipated across end-to-end conversion and distribution stages. Pie charts illustrate the breakdown of VR conduction and switching losses for the $A_1$-DVPD architecture.  %The power is assumed to be delivered to platform at 48 V. 
%The total power loss is shown as per cent of the total power load of 1\,kW.
}
\label{fig:comparison}
\end{figure}

To evaluate thermal feasibility, the optimized DVPD geometry and component-level loss distributions generated by the proposed framework are coupled to the thermal-fluidic solver of~\cite{choi2025self,choi2025automated}. In the example implementation of~\cite{choi2025self,choi2025automated}, substrate-embedded micro-channel cooling is employed, with cooling channels positioned adjacent to the GaN switch layer (PL$_2$), as illustrated in Figs.~\ref{fig:DVPD} and \ref{fig:VR-cell}. For the evaluated 1 kW $A_1$ configuration, the thermal analysis indicates a coolant flow rate of 1.6 g/s, limiting the maximum junction temperature to approximately 84°C. At thermal steady state, the total system loss converges to 210 W, corresponding to an end-to-end efficiency of 82.6\%. A detailed analysis of the DVPD thermal behavior, including spatial temperature distributions, temperature-dependent loss convergence, and cooling optimization, is provided in~\cite{choi2025self}.

\subsection{Scalability}

The developed DVPD design framework is inherently scalable to the multi-kilowatt demands of wafer-scale systems such as Cerebras~\cite{khairy2020} and the heterogeneous wafer-scale platforms outlined in the HIR roadmap~\cite{HIR2024}. For example, scaling from the 1\,kW reference case to a 25\,kW system at 2\,A/mm\textsuperscript{2} requires approximately 1,750 VRs, distributed across chiplets according to their thermal design power (TDP). Since the power handled per VR remains constant, the conversion efficiency is largely preserved as load power increases, enabling the DVPD system to maintain uniform performance across a wide power range, as illustrated in Fig.~\ref{fig:Power loss scalability}.  
The breakdown of power losses highlights two key scalability trends. First, conversion losses remain nearly constant with load power, as each VR processes the same power regardless of system scale. Second, supply-to-VR routing losses increase linearly with system power, since more current must be delivered from the system edges. This effect is strongly dependent on the input voltage: with 48\,V supply ($A_1$), the RMS current delivered to the VR layer is two times lower than with 12\,V supply ($A_3$), reducing distribution losses by approximately 4$\times$. As a result, high-voltage delivery ($A_1$) is preferable for wafer-scale systems exceeding 20\,kW, where lateral routing distances and currents dominate overall loss.

\begin{figure}[t!]
\centerline{\includegraphics[width=0.5\textwidth]{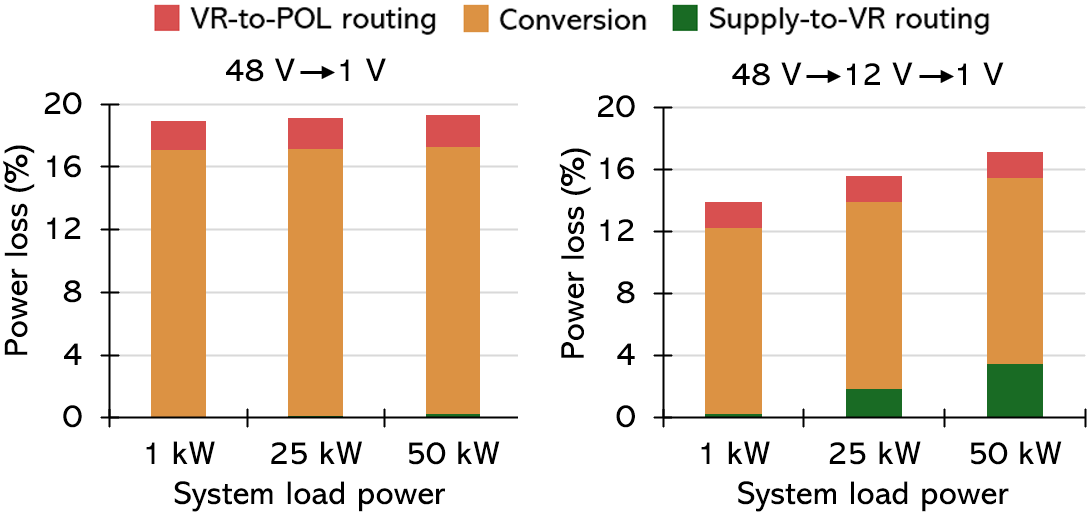}}
\caption{Percent of power loss in a DVPD system using architectures $A_1$ (48V/1V) and $A_3$ (48V/12V/1V), delivering a load power ranging between 1-50\,kW at 2\,A/mm\textsuperscript{2}.}
\label{fig:Power loss scalability}
%\vspace{-15pt}
\end{figure}

The maximum current density demonstrated at efficiencies above 80\% in this work is 2 A/mm², corresponding to the lower bound of the target 2–4 A/mm² range. The proposed framework can explore lower and higher current-density operating points by modifying design variables such as the number of vertically stacked layers, inductor density, and switching frequency. However, these choices introduce practical tradeoffs: higher density generally increases conduction, switching, and magnetic losses, thereby reducing efficiency and increasing local heat generation. As a result, the attainable current density is ultimately bounded not only by electrical constraints, but also by thermal feasibility, since localized hot spots can exceed 100 °C.  %\hl{Although the full self-consistent electrothermal co-design loop is addressed in the companion work~[39], the present framework incorporates temperature-dependent device characteristics and loss models, with thermal feasibility maintained through micro-channel flow rate adjustment, providing a basis for identifying density--efficiency tradeoffs across VPD systems}~\cite{choi2025self,choi2025automated}. 

%Although detailed thermal modeling is beyond the scope of this work, the proposed framework provides a basis for identifying density–efficiency tradeoffs, extending recent thermal approaches for VPD systems .

These findings demonstrate that the proposed DVPD framework not only preserves efficiency while scaling to tens of kilowatts, but also provides a systematic means to evaluate architectural tradeoffs and identify the most viable configurations. Future advances in inductor and capacitor technology will be essential to fully realize these scalability benefits.

\vspace{-5pt}
\section{Conclusion}\label{sec:Conclusion}

This work presents a comprehensive design and optimization framework for distributed vertical power delivery (DVPD) in advanced HPC and AI platforms. The framework integrates compact analytical models with automated floorplanning to co-optimize switching frequency, modular inductor scaling, interconnect allocation, and regulator placement, enabling systematic exploration of performance--density tradeoffs.

Several key findings emerge from this study. A methodology is established for selecting the minimum VR switching frequency that maintains ripple constraints across diverse topologies and packaging technologies. A modular inductor approach is introduced that maximizes current density and area utilization through scalable parallel integration. Analysis of vertical interconnects further reveals that required footprint scales with RMS rather than average current, providing guidance for balancing intermediate voltage selection against interconnect demand. In addition, floorplanning studies show that balanced cell geometries (aspect ratio $\approx 1$) reduce routing losses and improve density, while automated placement exposes spatial loss variation across the system.

System-level evaluation confirms that DVPD achieves substantially higher efficiency than conventional lateral or lumped vertical delivery. In particular, the evaluated 48~V/1~V DVPD system with 70 distributed regulators demonstrates 84\% end-to-end efficiency, compared to less than 70\% efficiency in representative centralized systems. Scalability analysis further shows that efficiency is preserved up to 50~kW operation at 2~A/mm$^2$, with high-voltage delivery providing the most favorable balance between conversion efficiency, interconnect area, and routing loss.

Overall, the proposed DVPD framework demonstrates the feasibility of delivering tens of kilowatts at high efficiency and density in next-generation heterogeneous HPC systems, while providing practical guidance for packaging integration, power converter design, and architectural tradeoffs.

\vspace{-10pt}

\bibliographystyle{IEEEtran}
\bibliography{./master_bib}

\end{document}